\documentclass[11pt,paper]{article}
\pdfoutput=1
\usepackage{jheppub}

\usepackage[usenames,dvipsnames,table]{xcolor}
\usepackage{graphicx,amsmath,amssymb,multirow,array,bm,mathrsfs}
\usepackage{epsf,amsfonts}
\usepackage[numbers,sort&compress]{natbib}
\usepackage{dsfont}
\usepackage{graphicx}
\usepackage{slashed}
\usepackage{float}
\usepackage{mathrsfs}%for scri
\usepackage{microtype}

\usepackage{pbox}
\usepackage{amsmath}
\usepackage{amsfonts}
\usepackage{pbox}
\usepackage{multirow}

\usepackage{pict2e}
\makeatletter
\DeclareRobustCommand{\loplus}{\mathbin{\mathpalette\dog@lsemi{+}}}
\newcommand{\dog@rsemi}[2]{\dog@semi{#1}{#2}{-90,90}}
\newcommand{\dog@lsemi}[2]{\dog@semi{#1}{#2}{270,90}}
\newcommand{\dog@semi}[3]{%
  \begingroup
  \sbox\z@{$\m@th#1#2$}%
  \setlength{\unitlength}{\dimexpr\ht\z@+\dp\z@\relax}%
  \makebox[\wd\z@]{\raisebox{-\dp\z@}{%
    \begin{picture}(1,1)
    \linethickness{\variable@rule{#1}}
    \roundcap
    \put(0.5,0.5){\makebox(0,0){\raisebox{\dp\z@}{$\m@th#1#2$}}}
    \put(0.5,0.5){\arc[#3]{0.5}}
    \end{picture}%
  }}%
  \endgroup
}
\newcommand{\variable@rule}[1]{%
  \fontdimen8
  \ifx#1\displaystyle\textfont3\else
    \ifx#1\textstyle\textfont3\else
      \ifx#1\scriptstyle\scriptfont3\else
        \scriptscriptfont3\relax
  \fi\fi\fi
}
\makeatother
\newcommand{\be}{\begin{equation}}
\newcommand{\ee}{\end{equation}}
\newcommand{\bea}{\begin{eqnarray}}
\newcommand{\eea}{\end{eqnarray}}

\newcommand{\half}{{1 \over 2}}

\newcommand*{\dd}{\mathop{}\!d}

\newcommand{\bes}{\begin{equation*}}
\newcommand{\ees}{\end{equation*}}
\newcommand{\beas}{\begin{eqnarray*}}
\newcommand{\eeas}{\end{eqnarray*}}

\newcommand{\ione}{{\tt p}}
\newcommand{\itwo}{{\tt q}}
\newcommand{\ithr}{{\tt r}}
\newcommand{\vp}{\varphi}

\newcommand{\cM}{\mathcal{M}}

\newcommand{\bmat}{\begin{bmatrix}}
\newcommand{\emat}{\end{bmatrix}}

\def\CL{{\cal L}}

\def\CO{{\cal O}}

\newcommand{\ben}{\begin{enumerate}}
\newcommand{\een}{\end{enumerate}}

\def\half{\frac{1}{2}}

\title{Warped Flatland}

\author[a]{St\'{e}phane Detournay,}
\author[a]{Wout Merbis,}
\author[b,c]{Gim Seng Ng}
\author[d]{and Raphaela Wutte}

\affiliation[a]{Physique Math\'{e}matique des Interactions Fondamentales, Universit\'{e} Libre de Bruxelles, and International Solvay Institutes, Campus Plaine - CP 231, 1050 Bruxelles, Belgium}
\affiliation[b]{School of Mathematics, Trinity College Dublin, Dublin 2, Ireland}
\affiliation[c]{Hamilton Mathematical Institute, Trinity College Dublin, Dublin 2, Ireland}
\affiliation[d]{Institute for Theoretical Physics, TU Wien, Wiedner Hauptstr. 8-10/136, A-1040 Vienna, Austria}

 \emailAdd{sdetourn@ulb.ac.be, wmerbis@ulb.ac.be, gng@math.tcd.ie, rwutte@hep.itp.tuwien.ac.at}

\abstract{
We study warped flat geometries in three-dimensional
topologically massive gravity.
They are quotients of global warped flat
spacetime, whose isometries are given by
the 2-dimensional centrally extended
Poincar\'{e} algebra. The latter can be obtained as a certain scaling limit of Warped AdS$_3$ space with a positive cosmological constant. We discuss the causal structure of the resulting spacetimes using projection diagrams.
We study their charges and thermodynamics, together with asymptotic Killing vectors preserving a consistent set of boundary conditions including them. The asymptotic symmetry group is given by a Warped CFT algebra, with a vanishing current level. A  generalization of the derivation of the Warped CFT Cardy formula applies in this case,  reproducing the entropy of the warped flat cosmological spacetimes.
}
\begin{document}
\noindent

\maketitle

\section{Introduction}

The identification of the microscopic origin of gravitational entropy is one of the most fertile puzzles in modern theoretical physics. Progress in this context has been especially significant when it involves the identification of black hole microstates, however, the same issue for cosmological horizons has remained more elusive so far. In the former case, an important breakthrough appeared in the seminal derivation of the microscopic entropy of five-dimensional extremal BPS black holes in the context of string theory by Strominger and Vafa \cite{Strominger:1996sh}. This result has pointed at the central role played by two-dimensional conformal symmetry as underlying the asymptotic or near-horizon dynamics of various classes of black holes \cite{Brown:1986nw,Strominger:1997eq,Cvetic:1997uw,Cvetic:1998xh}. This importance has become even more ubiquitous in recent years, when it was observed that 2d conformal symmetry could pertain to the description of generic classes of black holes, including real-world ones \cite{Carlip:1994gy,Carlip:1995cd,Guica:2008mu,Castro:2010fd,Carlip:2012ff,Kapec:2016aqd,Haco:2018ske,Ball:2019atb}. In addition, beautiful connections have recently been established between the asymptotic symmetries group of 4d Minkowski space, the BMS$_4$ group \cite{Bondi:1962px,Sachs:1962wk,Sachs:1962,Barnich:2010eb} and soft gravitons theorems \cite{Weinberg:1964ew,Weinberg:1965nx,Strominger:2013jfa,He:2014laa}, suggesting that asymptotically flat quantum gravity in (3+1) dimensions exhibits a 2d conformal symmetry \cite{Cachazo:2014fwa,Kapec:2014opa,Cheung:2016iub,Pasterski:2017kqt,Donnay:2018neh}. In the case of cosmological horizons, important advances could be made in the context of the dS/CFT correspondence \cite{Strominger:2001pn,Anninos:2011ui}, also featuring 2d conformal symmetry but many questions are left open \cite{Donnay:2019zif}.

For definiteness and simplicity, we will focus here on gravity in (2+1) dimensions. Its simplest occurrence, pure Einstein-Hilbert gravity, has proven to be a very rich toy model to address numerous questions, providing a simple setup for the gauge/gravity correspondence \cite{Deser:1984dr,Jackiw:1985je,Brown:1986nw,Achucarro:1986vz,Witten:1988hc,Verlinde:1989ua,Carlip:1991zm,Carlip:1995zj,Coussaert:1995zp}. The archetypical scenario consists of AdS$_3$ gravity with Brown-Henneaux boundary conditions, in which case the corresponding phase space exhibits a two-dimensional conformal symmetry with specific central extensions \cite{Brown:1986nw} and contains black holes solutions (the BTZ black holes \cite{Banados:1992wn,Banados:1992gq}); the entropy of which can be accounted for by the Cardy formula \cite{Strominger:1997eq}. This is summarized in the first column of Table \ref{cardytable}. 
It is worth mentioning that even in the canonical setup of AdS$_3$ gravity, a wealth of alternative boundary conditions has appeared over the years. These could either consist out of a relaxation of the original fall-offs or the presence of higher-order equations of motion or matter fields, resulting in either the same asymptotic symmetry algebra (ASA) \cite{Henneaux:2009pw,Henneaux:2010fy,Grumiller:2008es,Oliva:2009ip,Skenderis:2009nt,Giribet:2009qz,Henneaux:2002wm,Henneaux:2004zi,Henneaux:2006hk}, or in boundary conditions with different ASAs \cite{Compere:2013bya,Troessaert:2013fma,Afshar:2016kjj,Grumiller:2016pqb,Zwikel:2016smm}.

Various new developments have appeared in the last decade exploring departures from the conformal comfort zone. Spaces with non-(A)dS asymptotics have started to draw attention, including Schr\"odinger or Lifshitz spacetimes relevant to AdS/CMT \cite{Balasubramanian:2008dm, Son:2008ye,Kachru:2008yh}, Warped (A)dS$_3$ spaces (W(A)dS$_3$) \cite{Compere:2007in,Compere:2008cv,Compere:2009zj,Anninos:2010pm,Henneaux:2011hv,Blagojevic:2009ek,Anninos:2009jt,Anninos:2011vd}, near-horizon geometries of non-extremal black holes \cite{Donnay:2015abr,Afshar:2015wjm,Afshar:2016wfy,Afshar:2016kjj,Aggarwal:2019iay}, or flat space \cite{Barnich:2006av,Barnich:2010eb,Detournay:2016sfv,Grumiller:2017sjh}. Central to these approaches is the determination of consistent boundary conditions --- which defines the phase space of the theory at hand --- and the symmetries preserving them (forming the ASA), hinting at the structure of the corresponding quantum Hilbert space, as in AdS$_3$/CFT$_2$. In particular, two scenarios paralleling AdS$_3$ gravity have appeared in the last years: flat space holography in 2+1 dimensions, and WAdS$_3$ gravity. In both cases, the ASA does not display conformal symmetry. In the former case it consists in the three-dimensional Bondi-Sachs-Metzner-Van der Burg algebra (BMS$_3$), in the latter in the semi-direct product of a Virasoro and a $\hat{u}(1)$ Kac-Moody algebra.

These observations have hinted at the existence of new classes of two-dimensional field theories (respectively dubbed \emph{BMS field theories} and \emph{Warped Conformal Field Theories} or \emph{WCFTs} \cite{Hofman:2011zj,Detournay:2012pc}) as being to Mink$_{2+1}$ and WAdS$_3$ spaces what two-dimensional CFTs are to AdS$_3$. Efforts have been devoted in recent years in defining and exploring the properties of such theories, as well as exploiting them in a holographic context. One key point is that these theories, besides displaying an infinite-dimensional symmetry, enjoy a version of modular invariance allowing to single out regimes in which the density of states is captured by Cardy-like formulas \cite{Barnich:2012xq,Bagchi:2012xr,Detournay:2012pc}. In WAdS$_3$ gravity, the Bekenstein-Hawking entropy of the so-called spacelike warped black holes \cite{Bouchareb:2007yx,Moussa:2003fc,Anninos:2008fx} could then be reproduced by the counting of an asymptotic growth of states in a WCFT \cite{Detournay:2012pc}. Three-dimensional pure gravity with a vanishing cosmological constant, on the other hand, notoriously does not contain black hole solutions\footnote{For interesting attempts in flat space higher-curvature gravity, see e.g. \cite{Oliva:2009ip}}, but the flat limit of BTZ black holes, called Flat Space Cosmologies (FSC) \cite{Cornalba:2002fi} do enjoy interesting thermal properties and are endowed with a cosmological horizon and entropy. The latter can be matched to a BMS-Cardy formula counting the growth of states in a BMS-field theory \cite{Barnich:2012xq,Bagchi:2012xr}. This is summarized in the second and third column of Table~\ref{cardytable}.

\begin{table}
	\centering
	\resizebox{\textwidth}{!}{
		\begin{tabular}{l l l l}
			\hline
			& AdS$_3$  & Warped AdS$_3$ & 3D Flat space \\ \hline \\
			ASA & $[L_n^\pm, L_m^\pm] = (n-m) L_{n+m}^\pm + \frac{c^\pm}{12} (n^3-n) \delta_{n+m, 0}$  &
			$ \left[ L_n, L_m \right] = (n-m) L_{n+m} + \frac{c}{12} (n^3-n)\delta_{n+m,0}$  &
			$ \left[ L_n, L_m \right] = (n-m) L_{n+m} + c_L (n^3-n)\delta_{n+m,0}   $
			\\
			& &$[L_n, P_m] =  - m P_{n+m}$ & $[L_n, M_m] =  (n-m) M_{n+m}+ c_M (n^3-n)\delta_{n+m,0}$\\
			& &$ [P_n, P_m] = \frac{k}{2} n \delta_{n+m} $ & $ [M_n, M_m] = 0$\\
			[1 cm]
			Global    & $L^\pm_{0, \pm 1} \sim sl(2,R) \oplus sl(2, R) \subset {\rm Vir} \oplus {\rm Vir} $ &
			$L^\pm_{0, \pm 1}, P_0 \sim sl(2,R) \oplus u(1) \subset {\rm Vir} \loplus \hat{u}(1)$  &
			$L_{0, \pm 1}, M_{0, \pm 1} \sim iso(2,1) \subset {\rm bms}_3$ \\
			Subalgebra & & &
			\\ [1 cm]
			
			Field Theory
			& $x^\pm \rightarrow f^\pm (x^\pm)$ &
			$x^- \rightarrow f(x^-)$
			
			&
			$x^- \rightarrow f(x^-)$
			\\
			Symmetries & & $x^+ \rightarrow x^+ + g(x^-)$ & $ x^+ \rightarrow f'(x^-) x^+ + g(x^-) $
			\\ [1 cm]
			
			Solutions  & BTZ Black holes \cite{Banados:1992gq} & Warped AdS$_3$ Black holes (WBH) \cite{Moussa:2003fc,Anninos:2008fx} & Flat space cosmologies (FSC) \cite{Cornalba:2002fi} \\
			of Interest & & \\ [1 cm]
			
			\multirow{2}{*}{$\begin{aligned}
				&\mathrm{Degeneracy} \\
				&\mathrm{of\, States} \end{aligned}$}  &  $S_{\rm CFT} = 4 \pi  \sqrt{- L_0^{+, \mathrm{vac}} L_0^+} + 4 \pi \sqrt{- L_0^{-, \mathrm{vac}} L_0^- }$
			& $S_{\rm WCFT} = - \frac{4 \pi i}{k}  P_0^{\mathrm{vac}} P_0 + 2 \pi \sqrt{\frac{c}{6} \left(L_0 - \frac{P_0^2}{k} \right)} $ &
			$S_{{\rm BMS}_3} = 2 \pi \left( L_0 \sqrt{\frac{c_M}{2 M_0}} +  c_L \sqrt{\frac{M_0}{2 c_M}} \right) $ \nonumber \\
			&$\qquad \enskip= 2 \pi \sqrt{\frac{c^+ L_0^+}{6}} + 2 \pi \sqrt{\frac{c^- L_0^-}{6}}$   &  &\nonumber
			\\ [1 cm]
			&  $S_{\rm BTZ} = S_{\rm CFT}$ & $S_{\rm  WBH} = S_{\rm WCFT}$ & $S_{\rm FSC} = S_{{\rm BMS}_3}$ \\ [1cm]
			\hline
		\end{tabular}
	}
	\caption{Three scenarios in 3d gravity}
	\label{cardytable}
\end{table}

These three scenarios do not stand on an equal footing. The first has a long history, starting in the mid-eighties both on its constitutive parts (AdS spaces and CFTs) and on their close relationship (with the original proposal of \cite{Maldacena:1997re,Witten:1998qj}, see \cite{Eberhardt:2019ywk} for recent advances in deriving the correspondence). In the two latter cases, the field theories are far less understood, but they have attracted considerable attention in recent years. Their intrinsic properties, as well as explicit realizations, have been studied from various perspectives \cite{Hofman:2011zj,Detournay:2012pc, Hofman:2014loa,Castro:2015uaa,Detournay:2015ysa,Castro:2015csg,Song:2016gtd,Song:2016pwx,Song:2017czq,Anninos:2017cnw,Bardeen:1999px,Dias:2007nj,Guica:2008mu,Castro:2009jf,Azeyanagi:2012zd,Anninos:2013nja,Anninos:2008qb,Compere:2013aya,Davison:2016ngz,Chaturvedi:2018uov,Afshar:2019tvp,Jensen:2017tnb,Barnich:2012rz,Grumiller:2019xna}. Besides the scenarios of Table \ref{cardytable}, other holographic realizations featuring non-conformal (and actually, non-Lorentzian) field theories exist of two types: (i) Field theories with Lifshitz-type or more general anisotropic scalings \cite{Gonzalez:2011nz,Chen:2019hbj,Fuentealba:2019oty}, (ii) Irrelevant deformations of 2d CFTs \cite{Smirnov:2016lqw,Cavaglia:2016oda,Guica:2017lia}, relevant to the holographic description of classes of three-dimensional black strings \cite{Apolo:2019zai,Giveon:2017nie,Hyun:1997jv}. 

In the present work, we will analyze a class of three-dimensional spacetimes that does not fit in any of the above scenarios. This is referred to as Warped Flat (WF) spaces \cite{Moussa:2008sj,Anninos:2009jt} and can be viewed as a fibration over two-dimensional flat space along a fiber coordinate spanning the real line. They can be obtained as a scaling limit of WAdS$_3$ or WdS$_3$ spacetimes, where a certain flat limit of the (A)dS$_2$ basis is taken. Global identifications of WF were discussed in \cite{Moussa:2008sj} in the context of Topologically Massive Gravity \cite{Deser:1981wh,Deser:1982vy} coupled to Maxwell theory with an electromagnetic Chern-Simons term. The so-called self-dual WF quotient appears as the near-horizon geometry at fixed polar angle of the ultra-cold limit of Kerr-dS black holes where the inner, outer and cosmological horizons coincide \cite{Anninos:2009yc}, much like self-dual spacelike WAdS$_3$ appears in the near-horizon limit of extremal Kerr \cite{Bardeen:1999px,Guica:2008mu}.
Another type of quotient was shown to result in a two-parameter family of spacetimes exhibiting a Killing horizon and claimed to describe causally regular black holes \cite{Moussa:2008sj,Anninos:2009jt}.
 They will be the subject of this work. 
 Contrary to claims in \cite{Moussa:2008sj,Anninos:2009jt} we will argue that these solutions represent cosmological spacetimes, whose horizon is endowed with a non-trivial entropy which,  when expressed in terms of the global charges of the solutions (denoted by $P_0$ and $L_0$), 
 takes the functional form
\begin{equation}\label{WFEntropy}
  S =  \alpha_1 \, P_0 + \alpha_2 \, \frac{L_0}{P_0},
\end{equation}
where $\alpha_{1,2}$ are constants (see (\ref{WFS}) for the precise expression). This dependence of the entropy on the charges of the solutions is not of any of the three types that can be seen in Table \ref{cardytable}. 
The goal of this paper will be to find a set of boundary conditions accommodating these quotients of Warped Flat space (whose geometric structure will be carefully analyzed), determine their symmetries, and analyze whether those could be used to reproduce the entropy (\ref{WFEntropy}).

The paper is organized as follows. In section \ref{sec:geometry}, we introduce various aspects of the geometries we are interested in, in particular their unusual isometries. We analyze their causal structure and conclude that they do not describe black holes, but rather cosmological spacetimes for a certain range of the parameters. In section \ref{TMGfirstorder}, we embed the solutions in a dynamical theory, which we choose to be Topologically Massive Gravity in Chern-Simons-like form \cite{Carlip:2008qh,Bergshoeff:2014bia,Merbis:2014vja}. We work out a set of on-shell boundary conditions including the geometries of interest and determine their asymptotic symmetries. We also compute the thermodynamic quantities of interest, such as mass, angular momentum, Hawking temperature, and Bekenstein-Hawking entropy. In section \ref{sec:WCFT} we turn to a field theory analysis of the symmetries found in the previous section. In particular, following the analysis of \cite{Detournay:2012pc,Afshar:2015wjm}, we derive a Cardy-type formula for a centerless WCFT, which we show matches the geometric entropy previously obtained. Appendix \ref{app:WFaslimits} discusses the limit from spacelike W(A)dS$_3$ to WF spacetimes. In \ref{wfqe} we discuss the warped flat limit of warped AdS$_3$ as a deformation of a Euclidean Kerr-de Sitter spacetime and find that the warped flat quotient can be understood as a deformation of a locally flat spacetime.
Appendix \ref{app:WAdS} provides an on-shell version of the WAdS$_3$ boundary conditions \cite{Compere:2009zj,Henneaux:2011hv} with a positive cosmological constant, in the spirit of the Ba\~nados metrics \cite{Banados:1998gg} for AdS$_3$ (see e.g. \cite{Sheikh-Jabbari:2016unm} and references therein). Appendix \ref{appdercardy} gives extra details on the warped conformal field theory derivation of the entropy.
Have fun reading!

\section{Geometry of Warped Flat Spacetimes}\label{sec:geometry}
In this section we consider the geometry of the warped flat spacetimes mentioned in the introduction. We start by discussing
warped flat space in subsection \ref{wfspace} and compute its global Killing vectors
and the finite coordinate transformations generated by them. Then, we introduce a 
particular quotient of warped flat space in subsection \ref{wfquotient}.
Depending on the parameters of the warped flat quotient, the spacetime
may contain closed timelike curves. 
We discuss the causal structure of both warped flat space and the warped flat quotient in section \ref{causalwf} 
using techniques developed in \cite{Chrusciel:2012gz}
and find that
both spacetimes are, in fact, not black holes.
In particular, the causal diagrams of warped flat space and the 
warped flat quotient in the case where no closed timelike curves are present
coincide with the one of Minkowski space.
We find that for the case where closed timelike curves are present, the causal
diagram for the warped flat quotient is the same as that of a flat space cosmology.

\subsection{Warped Flat Space} 
\label{wfspace}
We consider the following three-dimensional spacetime \cite{Moussa:2008sj,Anninos:2009jt}:
\be\label{eqmetricvacuum}
\frac{ds^2}{\ell^2}=
dx^2-d\tau^2
+12\left(dy+ x d\tau \right)^2
=dx^2+12 dy^2 + 24 x d\tau dy
+\left(12x^2 - 1\right)d\tau^2,
\ee 
which was dubbed warped flat spacetime in \cite{Anninos:2009jt}.
Here, the coordinates $\tau,x,y$ range over the real numbers. 
The metric \eqref{eqmetricvacuum} is smooth
as far as curvature invariants are concerned, which read $R=6/\ell^2$,  $R_{\mu\nu}R^{\mu\nu} = 108/\ell^4$. 
The Cotton tensor (which is analogous to the Weyl tensor in higher dimensions) reads:
\be
C_{\mu\nu}dx^\mu dx^\nu
= \frac{12\sqrt{3}}{\ell}\left[
\left(24x^2 +1 \right) d\tau^2
+48 x d\tau dy
-dx^2
+24 dy^2
\right]\,,
\ee and $C_{\mu\nu}C^{\mu\nu}=2592/\ell^6$
\footnote{For computing the Cotton tensor
we have chosen the following orientation of the epsilon tensor $\epsilon^{\tau x y}=1/(\sqrt{- \mathrm{det}g}).$}
The metric \eqref{eqmetricvacuum} can be obtained as a limit of global warped Anti-de Sitter space (WAdS) or
warped de Sitter space (WdS) (both with positive cosmological constant) \cite{Anninos:2009jt}. 
Detailed discussions of the appropriate limits are provided in Appendix \ref{app:WFaslimits}.

The inverse metric in coordinates $(\tau,x,y)$ reads
\be\label{inmet}
\ell^2 g^{\mu\nu}=\left(
\begin{array}{ccc}
 -1 & 0 & x \\
 0 & 1 & 0 \\
 x &0 & \frac{1}{12}\left(1-12x^2\right)
\end{array}
\right).
\ee From \eqref{inmet} it can be seen that the normal vector to constant 
$\tau$ surfaces is always timelike while the one to constant $x$ surfaces is spacelike. However, for constant $y$ surfaces, $12 \ell^2 n_\mu n^\mu = 1-12x^2$; so it is a spacelike surface for $|x|>1/\sqrt{12}$, while for $|x|<1/\sqrt{12}$ it is a timelike surface. The $x=\pm1/\sqrt{12}$ surfaces are null surfaces.

The exact isometries are generated by the four Killing vectors
\bea
\label{isom}
I_0 &=& -2 \partial_y \,, \nonumber\\
a_\pm &=&
\partial_\tau \mp \partial_x \pm \tau \partial_y \,,
\\ \nonumber
H &=& -\tau \partial_x - x\partial_\tau+\half\left(x^2+\tau^2\right) \partial_y\,,
\eea
satisfying the following algebra:
\be
\label{isomalgebra}
[a_+, a_-] = I_0,~~
[H,a_\pm]= \mp a_\pm.
\ee where $I_0$ commutes with all other generators.
The algebra is precisely the one of the Hamiltonian, annihilation and creation operator  
of a harmonic
oscillator in quantum mechanics, where $I_0$ is a c-number.
 The algebra   \eqref{isomalgebra} is known under the name
 $P_2^c$, as it is the 2-dimensional centrally extended
 Poincar\'{e} algebra. We may bring the commutation relations
 into a well-known form through the following change of
 basis $\sqrt{2} a_+ = P_1 + P_0$,
 $\sqrt{2} a_- = P_1 - P_0$ leading to
 \be
 \label{p2c}
 [P_0, P_1] = I_0,~~
 [H,P_1]= - P_0, ~~
 [H, P_0] = - P_1\,.
 \ee
 Here, $H$ denotes the boost, $P_0$ and $P_1$ are the
 2-dimensional translations and $I_0$ denotes the central
 extension. $P_2^c$ can also be obtained as an \.In\"on\"u-Wigner contraction of the $sl(2,R) \loplus u(1)$ algebra, as shown in App. A.
 Finally, it can viewed as a global subalgebra of $Vir \loplus  \; \hat{u}(1)$ with $L_0 = H$, $P_0 = I_0$, $L_{+1} = a_+$ and $P_{-1} = a_-$.

 We see that some of these isometries have a natural geometric interpretation.
 For $I_0$ and $a_+ + a_- =2\partial_\tau$, the finite coordinate transformations are translations in $y$ and $\tau$. For $- \frac12 (a_+-a_-)= -\tau\partial_y+\partial_x$, we have the simultaneous transformation
\bea
x'=x+C\,, \qquad
y'=y-C \tau\,,\qquad
\tau'=\tau\,.
\eea 
Here, $C$ is an arbitrary constant.
Finally the finite transformation generated by $H$ is the most complicated one:
\bea
\tau'&=&\tau \cosh C- x \sinh C \,, \nonumber \\
x'&=&- \tau\sinh C + x \cosh C \,, \nonumber \\
y'
&=&
y+\frac{1}{2} \sinh (C) \left[\cosh (C) \left(\tau^2+x^2\right)-2 \tau x \sinh (C)\right]\,.
\eea We can see that the $\tau'$ and $x'$ transformations are simply boost transformation, while the $y'$ transformation is non-trivial and 
does not allow for a simple geometric interpretation.

\subsection{Quotienting Warped Flat Space}
\label{wfquotient}

Following \cite{Moussa:2008sj}, we start with the warped flat spacetime \eqref{eqmetricvacuum}
and consider the region $x^2-\tau^2>0, x>0$.
 We perform the coordinate transformation
\bea
\label{coordinatetraf}
x&=& \sqrt{\frac{\rho}{6\xi}} \cosh{(12 \xi \varphi)} \,, \nonumber\\
\tau&=& \sqrt{\frac{\rho}{6\xi}} \sinh{(12 \xi \varphi)} \,,  \\ \nonumber
y&=&u +(\xi+\omega)\varphi - \frac{\rho}{24 \xi} \sinh{(24 \xi \varphi)} \,,
\eea
where $\xi$ and $\omega$ are two real constants and $\rho/\xi > 0 $.
 We obtain
\bea
\label{wfbhxi}
\frac{ds^2}{\ell^2} &=& \frac{d\rho^2}{24 \xi \rho}
+12 du^2
+24 (\rho+\omega+\xi) d\varphi du
+12\left[\left(\rho+\omega\right)^2+\xi(\xi+2\omega) \right] d\varphi^2 \nonumber\\
&=&
\frac{d\rho^2}{24 \xi \rho}
-24 \xi \rho d\varphi^2
+12\left[du+(\rho+\xi+\omega)d\varphi \right]^2\,,
\eea which upon identification of $(u, \rho, \varphi) \sim (u, \rho, \varphi + 2 \pi)$ gives the warped flat quotient. This amounts to perform discrete identifications in the global warped flat geometry (\ref{eqmetricvacuum}) along orbits of the Killing vector
\be \label{IKV}
 \partial_\varphi = -12 \xi H - \frac{\xi + \omega}{2} I_0.
\ee
Here, $u$ runs from $-\infty$ to $\infty$. 
Depending on the sign of $\xi$, $\rho$ runs from $-\infty$ to 0 or from 0 to $\infty$.
If we send $\rho \rightarrow - \rho, \xi \rightarrow - \xi,  \omega \rightarrow - \omega$ and $\varphi \rightarrow - \varphi$ the metric, the identifications, and the orientation stay the same. 
Thus, $(\xi, \omega)$ and $(-\xi, -\omega)$ describe spacetimes which are isometric with the same orientation. 
Hence, in the following we restrict to $\xi >0$ without loss of generality.

The inverse metric (in $(u, \rho,\varphi)$ coordinates) reads
\begin{equation}\ell^2 g^{\mu\nu}=\left(
\begin{array}{ccc}
    -\frac{(\rho+\omega)^2+\xi(\xi+2\omega)}{24 \xi \rho} & 0 & \frac{\rho+\xi+\omega}{24 \xi \rho} \\
  & 24\xi \rho &  0\\
  & & -\frac{1}{24 \xi \rho}
\end{array}
\right).
\end{equation}
The global Killing vectors $H$ and $I_0$ as well as the local Killing vectors $a_\pm$ are given by
\begin{subequations}
\bea\label{globalKV}
H&=& \frac{1}{12\xi}\left((\xi+\omega) \partial_u -\partial_\varphi\right),\quad
I_0 = -2 \partial_u, \\
a_\pm &=& \frac{e^{\pm 12 \xi \varphi}}{2 \sqrt{6 \xi \rho}}\left[
\partial_\varphi
+\left(\rho-\xi-\omega\right) \partial_u
\mp 24 \xi \rho \partial_\rho \label{localKV}
\right],
\eea 
\end{subequations}
satisfying $[a_+,a_-]= I_0,~[H,a_{\pm}]= \mp a_{\pm}$. 

The inverse transformation of \eqref{coordinatetraf} reads
\begin{align}
	\label{inversetraf}
    \rho(x, \tau) &= 6 \xi (x^2- \tau^2) \,, \nonumber\\
u(\tau, x, y) &= y+ \frac{1}{4} \left(x^2-\tau ^2\right) \sinh \left(2 \mathrm{arctanh}\left(\frac{\tau
	}{x}\right)\right)-\frac{(\xi +\omega ) \mathrm{arctanh} \left(\frac{\tau
	}{x}\right)}{12 \xi }\,,\\ \,
	\varphi(\tau, x) &= \frac{\mathrm{arctanh}\left(\frac{\tau}{x }\right)}{12 \xi}\,.\nonumber
\end{align}
For the parameter range $\omega \leq - \xi/2$ the metric 
component $g_{\varphi \varphi}$ becomes negative and  closed timelike curves occur in the region $\rho > 0$ between 
$\rho_1 = - \omega - \sqrt{- \xi (\xi + 2 \omega)} $ and 
  $ \rho_2 =  - \omega + \sqrt{- \xi (\xi + 2 \omega)} $.
In the following we will differentiate between the two cases of interest:
  \begin{enumerate}
  \item
   For $\omega > -\xi/2$: no closed timelike curves appear
    \item  
     For $\omega \leq -\xi/2$: closed timelike curves appear in the region $ - \omega - \sqrt{- \xi (\xi + 2 \omega)} < \rho <  - \omega + \sqrt{- \xi (\xi + 2 \omega)} $
\end{enumerate}
In the case  $\xi+\omega = 0$ closed timelike curves start to occur at 	$\rho = 0$, which is why we will often restrict to $\xi+\omega >0$ in the following.
The surface $\rho = 0$ is a Killing horizon of the following Killing vector
\be
K^\mu \partial_\mu = \partial_u -\frac{1}{\xi+\omega} \partial_\varphi \,
\ee
if $\xi + \omega \neq 0$.
 For $\xi + \omega = 0$ the Killing horizon is generated by the Killing vector $\partial_\varphi$.

\subsection{Causal Structures}
\label{causalwf}
In this section we discuss the causal structure of warped flat space \eqref{eqmetricvacuum} and the warped flat quotient
\eqref{wfbhxi}, using methods developed in \cite{Chrusciel:2012gz}.
In \cite{Chrusciel:2012gz} a new class of two-dimensional diagrams, 
the so-called \emph{projection diagrams}, were introduced as a tool to visualize the global structure of spacetimes.
These diagrams can be used to depict non-spherically symmetric or non-block diagonal metrics with two-dimensional diagrams, using 
a two-dimensional auxiliary metric constructed out of the spacetime.
For this one uses a map $\pi$ which maps (a subset of) the spacetime $U$ to 
(a subset of) $R^{1,1}$. This map is constructed such that every timelike curve in $U$
gets projected to a timelike curve in $R^{1,1}$ and each timelike curve in $\pi(U)$ is the projection of a timelike curve in the original spacetime.
This way causal relations in $\pi(U)$ reflect causal relations in $U$, see
\cite{Chrusciel:2012gz} for a precise definition.

In regions where closed timelike curves appear, causality is not represented 
in any useful way in the projection diagram. For this reason these regions are removed from
the diagram.

\subsubsection{Warped Flat Space}
First, we consider warped flat space \eqref{eqmetricvacuum}.
The map $\pi$ from the definition of the projection diagram is given by the projection $(\tau, x, y) \mapsto (\tau, x)$.
The auxiliary metric reads
\begin{equation}
    \label{auxwf}
	\gamma_{\mu \nu}d x^\mu d x^\nu := \ell^2 ( \dd x^2 - \dd \tau^2)\,, 
\end{equation}
which is two-dimensional Minkowski space $R^{1,1}$, whose conformal
compactification and conformal boundaries are well-known.

We now want to answer the question whether the geometry \eqref{eqmetricvacuum}
possesses a non-zero black hole region. 
Our notion of future asymptotic infinity $\mathscr{I}^+$ of \eqref{auxwf} will be defined with respect to 
 the conformal boundary of the two-dimensional metric. 
The two-dimensional spacetime \eqref{auxwf} does not have
 a black hole region, as the whole spacetime lies in the causal past of $\mathscr{I}^+$, i.e.\ $R^{1,1} - J^-(\mathscr{I}^+) = \emptyset$.

 This can be seen explicitly by considering the family of null curves $(\tau(s), x(s)) = (\tau_0 + s,x_0+ s)$ going from each point in spacetime all the way to $\mathscr{I}^+$.
 These curves can be lifted to null curves in the three-dimensional spacetime going through every point:
 $(\tau(s), x(s), y(s)) = (\tau_0 + s,x_0+ s, y_0 - x_0 s + \frac{s^2}{2})$.
 This shows that there is no black hole region in our three-dimensional spacetime
 \footnote{We thank Piotr Chru\'{s}ciel for pointing this out to us.}.

\subsubsection{Warped Flat Quotient}
\begin{figure}
	\centering
  \includegraphics{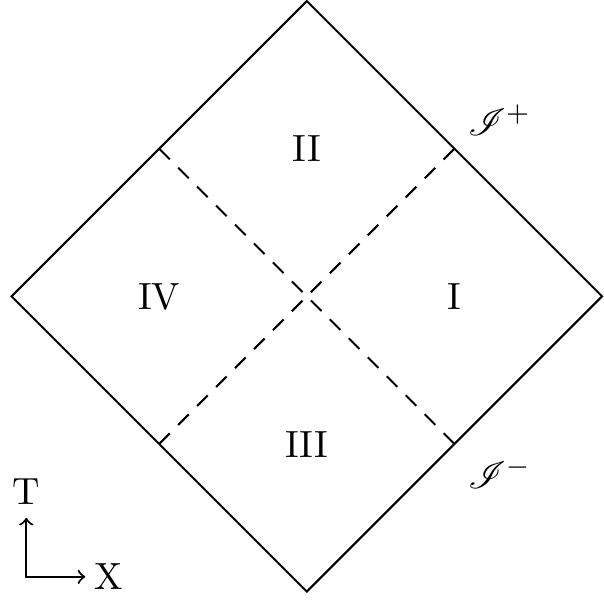}
    \caption{Here, we depict the split of the spacetime into the four sectors in the compact coordinates $T$ and $X$, which
    will be introduced below. Each coordinate patch $u, \rho, \varphi$ covers one such sector. The dashed lines denote the location of the Killing horizon. They intersect at the point $T = X = 0$.}
	\label{pic4sec}
\end{figure}

Next, we consider the warped flat quotient. Here we will differentiate between two cases, the case where CTCs are present 
and the case without CTCs.
As the coordinate system $(u, \rho, \varphi)$ used in the 
previous section cannot be extended beyond $\rho = 0$, we will work with the coordinates $(\tau, x, y)$
which are everywhere well-defined.
We split our spacetime into four sectors:
\begin{itemize}
\item[I]  $x^2 - \tau^2 > 0$, $\tau -x <0$
\item[II] $x^2 - \tau^2 < 0$, $\tau -x >0$
\item[III] $x^2 - \tau^2 < 0$, $\tau -x <0$
\item[IV] $x^2 - \tau^2 > 0$, $\tau -x > 0$
\end{itemize}
This split is depicted in figure \ref{pic4sec}.
The coordinate transformation 
in sector I has already been discussed in the previous section (see equation \eqref{coordinatetraf}).
In the sectors
$II$, $III$, $IV$ we introduce new coordinates $u$, $\rho$ and $\varphi$ as follows
\begin{subequations}
\label{wftowfbh}
\begin{align}
	II : x &=  \sqrt{-\frac{\rho }{6 \xi }} \sinh (12 \xi  \varphi ), \\
  \tau &=  \sqrt{-\frac{\rho }{6 \xi }} \cosh (12 \xi  \varphi),\\
    y &= \frac{\rho  \sinh (24 \xi  \varphi )}{24 \xi }+\varphi  (\xi
   +\omega )+u  \,, 
 \end{align}
 \begin{align}
   III : x &= - \sqrt{-\frac{\rho }{6 \xi }} \sinh (12 \xi  \varphi ), \\
  \tau &= - \sqrt{-\frac{\rho }{6 \xi }} \cosh (12 \xi  \varphi),\\
    y &= \frac{\rho  \sinh (24 \xi  \varphi )}{24 \xi }+\varphi  (\xi
   +\omega )+u \,,\\
IV : x &= -\sqrt{\frac{\rho }{6 \xi }} \cosh (12 \xi  \varphi
),\\
\tau &= -\sqrt{\frac{\rho }{6 \xi }} \sinh (12 \xi  \varphi
), \\
y &= -\frac{\rho  \sinh (24 \xi  \varphi )}{24 \xi }+\varphi  (\xi
+\omega )+u\,.
\end{align}
\end{subequations}
This leads to the metric \eqref{wfbhxi} in each sector, which upon identification of $(u, \rho, \varphi) \sim (u, \rho, \varphi + 2 \pi)$ gives the warped flat quotient.
Here, $u$ runs from $-\infty$ to $\infty$. 
Depending on the sector $\rho$ runs from $-\infty$ to 0 or from 0 to $\infty$.
The inverse transformation of \eqref{wftowfbh} reads
\begin{subequations}
	\label{rhototau}
\begin{equation}
    \rho = 6 \xi (x^2- \tau^2)\,,
\end{equation}
\begin{align}
x^2- \tau^2 < 0: u &=	y-\frac{1}{4} \left(x^2-\tau ^2\right) \sinh \left(2 \mathrm{arctanh} \left(\frac{x}{\tau }\right)\right)
	-\frac{(\xi +\omega ) \mathrm{arctanh}\left(\frac{x}{\tau }\right)}{12 \xi },\\
	\varphi &= \frac{\mathrm{arctanh}\left(\frac{x}{\tau }\right)}{12 \xi}   ,
\end{align}
\begin{align}
x^2- \tau^2 > 0: u &= y+ \frac{1}{4} \left(x^2-\tau ^2\right) \sinh \left(2 \mathrm{arctanh}\left(\frac{\tau
	}{x}\right)\right)-\frac{(\xi +\omega ) \mathrm{arctanh} \left(\frac{\tau
	}{x}\right)}{12 \xi },\\ \,
	\varphi &= \frac{\mathrm{arctanh}\left(\frac{\tau}{x }\right)}{12 \xi}\,.
\end{align}
\end{subequations}

The starting point of the construction is to write the metric \eqref{wfbhxi} in the following form:
\begin{align}
\frac{g_{\mu \nu} d x^\mu d x^\nu}{\ell^2} &=-\frac{24  \xi \rho}{\xi^2+2\xi \omega+(\rho+\omega)^2} du^2 + \frac{d\rho^2}{24 \xi \rho}\nonumber \\
&\quad+12 \left((\rho+\omega)^2  + 2\xi\omega +\xi^2\right) \left(d\varphi+  \frac{\rho + \xi +\omega}{\xi^2+2\xi \omega + (\rho+\omega)^2} du\right)^2
.
\end{align} 
We see that the last term is positive everywhere, except for in the region where closed timelike curves are present
(compare with \eqref{wfbhxi}). We discuss the case without closed timelike
curves first for which the last term is manifestly positive.
We project in such a way that the auxiliary metric $\gamma_{\mu \nu}$ reads
\begin{align}
    \label{auxgamma2}
    \gamma_{\mu \nu}d x^\mu d x^\nu := -\frac{24  \xi \rho}{\xi^2+2\xi \omega+(\rho+\omega)^2} du^2 + \frac{d\rho^2}{24 \xi \rho}\,.
\end{align}
Then we perform the following coordinate transformation in each of the four sectors
\begin{subequations}
    \label{UV}
    \begin{align}
 &I :& V &=  e^{c f(\rho) - c u}, & U &= - e^{c f(\rho) + c u}\,, \\
 &II :& V &=  e^{c f(\rho) - c u}, &  U &=  e^{c f(\rho) + c u}\,, \\
 &III :& V &= - e^{c f(\rho) - c u}, & U &= - e^{c f(\rho) + c u}\,, \\
 &IV :& V &= - e^{c f(\rho) - c u}, &  U &=  e^{c f(\rho) + c u}\,, 
    \end{align}
    \end{subequations}
where $c = \frac{12 \xi}{\xi+\omega}$
and $f(\rho)$ is the solution to the differential equation
\begin{equation}
	\label{diffeq}
	f'(\rho) = \frac{\sqrt{\xi^2 + 2 \xi \omega + (\rho + \omega)^2}}{24 \xi \rho}\,.
\end{equation}
The solution satisfies that $f(\rho= \pm \infty) = \infty$ and $f(\rho = 0) = - \infty$. 
 The coordinates V, U both run from $(-\infty, \infty)$.
We introduce two more coordinates
\begin{equation}
	V =\tan \left(\frac{X+T}{2} \right)\,, \quad U =\tan \left(\frac{T-X}{2} \right)\,,
\end{equation}
so that $(T-X)/2 \in (-\pi/2, \pi/2)$ and $(T+X)/2 \in (-\pi/2, \pi/2)$.
Now we can rewrite $\gamma_{\mu \nu}$ as 
\begin{align}
	\gamma_{\mu \nu}d x^\mu d x^\nu
	=& - \mathrm{sgn}(\rho) \frac{e^{-2 c f(\rho)} \rho (\xi+\omega)^2 }{6 \xi (\xi^2 + 2 \xi \omega + (\rho+ \omega)^2)}  \ell^2  \dd U \dd V \nonumber \\
	=& \frac{1}{\Omega^2} \ell^2 (- \dd T^2 + \dd X^2)\,.
\end{align}

\begin{figure}
	\centering
	\includegraphics{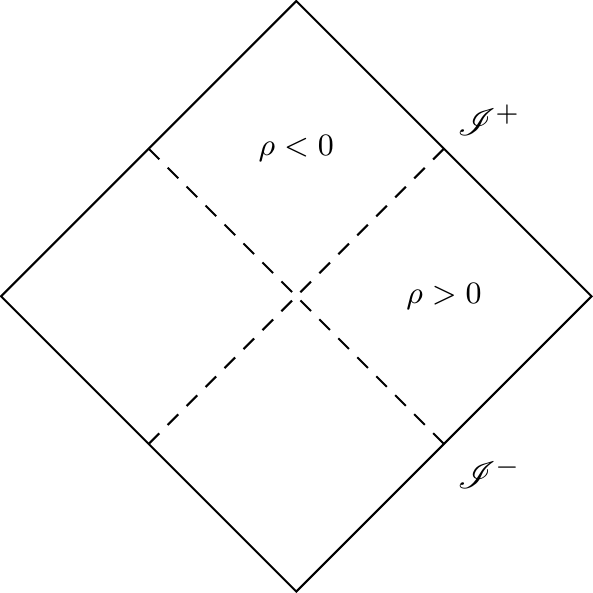}
	\caption{Projection diagram of the warped flat quotient in the case where
		no closed timelike curves are present ($\omega > - \xi/2$). The upper and lower sectors correspond to 
		$\rho < 0$, while the left and right sectors correspond to $\rho >0$. 
		The dashed lines that divide these sectors are the $\rho = 0$ lines.
		The four vertices correspond to $\rho \rightarrow \pm \infty, u = \mathrm{const.}$ }
	\label{fig3}
\end{figure}

\begin{figure}[t]
	\centering
	\includegraphics{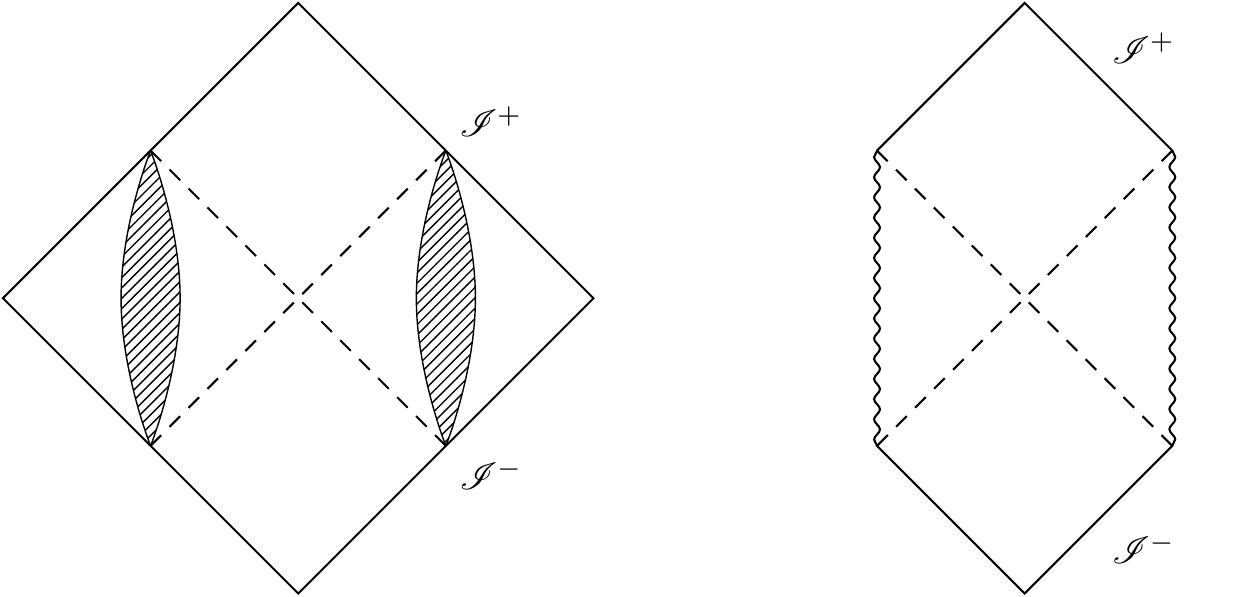}
	\caption{Projection diagram of the warped flat quotient in the case where
		closed timelike curves are present $\omega \leq - \xi/2$. The closed timelike curves appear in the shaded region in the left picture: $ - \omega - \sqrt{- \xi (\xi + 2 \omega)} < \rho <  - \omega + \sqrt{- \xi (\xi + 2 \omega)} $.
		In the right picture we have cut off the spacetime
		at the place where the closed timelike curves appear. The vertical, wiggly line
		is the singularity.
	}
	\label{fig2}
\end{figure}

The projection $\pi$ is then defined as the map $(\tau, x, y) \rightarrow (X(\tau,x, y), T(\tau, x, y))$.
The map is differentiable everywhere.
The conformal factor reads\footnote{Here we have introduced the conformal factor in the standard way $\tilde{g}_{\mu \nu} = \Omega^2 g_{\mu \nu}$, with $\tilde{g}_{\mu \nu}$ being the unphysical metric.}
\begin{equation}
	\Omega^2 = 4 \,\mathrm{sgn}(\rho) \frac{6 \xi e^{2 c f(\rho)} (\xi^2 + 2 \xi \omega + (\rho+ \omega)^2)  }{\rho (\xi+\omega)^2} \frac{1}{(1+ U^2(\rho,u))(1+V^2(\rho,u))}\,.
\end{equation}
The conformal factor goes to 0 as $\rho, u$ go to $\pm$ infinity, is regular at $\rho = 0$ and is positive everywhere.
The projection diagram of the warped flat quotient in the case where no closed
timelike curves occur looks like the one of two-dimensional Minkowski space and is 
depicted in figure \ref{fig3}.
The above derivation is valid for the case where no closed timelike curves appear.
In the case where closed timelike curves appear, the construction is valid everywhere except for 
in the region $ - \omega - \sqrt{- \xi (\xi + 2 \omega)} < \rho <  - \omega + \sqrt{- \xi (\xi + 2 \omega)} $
which must be excised from the diagram. We thus cut off our spacetime at $\rho = - \omega - \sqrt{- \xi (\xi + 2 \omega)}$. The resulting diagram is depicted in 
figure \ref{fig2}. 
This projection diagram looks like the Penrose diagram of flat space cosmologies \cite{Cornalba:2002fi}.

We now want to answer the question whether the geometry \eqref{wfbhxi}
possesses a non-zero black hole region. 
As before, our notion of future asymptotic infinity $\mathscr{I}^+$ of \eqref{auxwf} will be defined with respect to 
 the conformal boundary of the two-dimensional metric. The two-dimensional 
 spacetime \eqref{auxgamma2} does not have a black hole region.
To show that the three-dimensional geometry does not possess a black hole region either
we proceed as follows. In each sector we consider null curves in the two-dimensional geometry
\eqref{auxgamma2}. These curves $(u(\rho), \rho)$ are solutions to the differential
equation
\begin{equation}
    \label{diffnull1}
   \left( \frac{ \partial u(\rho)}{\partial \rho} \right)^2 = \left( \frac{\xi^2 +2 \xi \omega + (\rho+ \omega)^2}{{(24 \xi \rho)}^2} \right)\,,
\end{equation}
and may be lifted to null curves $(u(\rho), \rho, \varphi(\rho))$ in the three-dimensional geometry provided that 
\begin{equation}
    \label{diffnull2}
    \frac{ \partial \varphi(\rho)}{\partial \rho}  = - \frac{\rho + \xi + \omega}{\xi^2 + 2 \xi\omega + {(\rho + \omega)}^2 } \frac{\partial u(\rho)}{\partial \rho}\,.
 \end{equation}
The differential equations can be solved to give two curves
emanating from every point in spacetime, except at $\rho = 0$, where the coordinate 
system breaks down.
Considering the curves in the global coordinate system $(\tau, x, y)$ (see
\eqref{coordinatetraf} and \eqref{wftowfbh}), we find that the coordinates $(\tau(\rho), x(\rho), y(\rho))$
are finite and continuous for any $\rho$ if one patches the curves in the
sectors I, II, III, IV together appropriately.
As there exist such null curves going through every point this shows there is no black hole region in our three-dimensional spacetime.

\subsubsection{Warped Flat Quotient: \texorpdfstring{$\xi+\omega = 0$}{xi + omega = 0}}

\begin{figure}[t]
	\centering
	\includegraphics{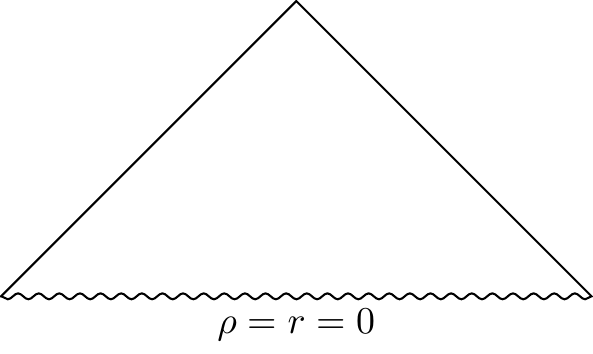}
	\caption{Projection diagram of the warped flat quotient in the case where $\xi + \omega = 0$. Due to the occurence of closed timelike curves the spacetime is cut off at $\rho = r= 0$, resulting in a singularity (wiggled line).
	}
	\label{figxiomega0}
\end{figure}

The causal analysis in the previous section holds true for all $\xi$, $\omega$ except for the case $\xi+\omega = 0$. Reconsidering \eqref{UV} for $\xi+\omega = 0$ we see that $c = \frac{12 \xi}{\xi+\omega}$ diverges if $\xi+\omega = 0$. 
The case $\xi + \omega = 0$ is special because -- as already briefly mentioned at the end of section 2.2 -- for this case closed timelike curves appear between the horizon $\rho = 0$ and $\rho = 2 \xi$. As causality is not represented in any useful way in this region, we must excise it from the 
diagram. We will thus cut off our spacetime at $\rho = 0$.
The metric \eqref{wfbhxi} for $\xi + \omega = 0$ reads 
\begin{equation}
	\frac{ds^2}{\ell^2}= \frac{d \rho^2}{24 \xi \rho} + \frac{24 \xi}{2 \xi - \rho} du^2 - \frac{12 \rho {(du+ (\rho - 2 \xi)d\varphi)}^2}{2 \xi - \rho}
\end{equation}
Performing the coordinate transformation  $\rho = - 6 r^2 \xi$ we obtain 
\begin{equation}
	\frac{ds^2}{\ell^2}= - dr^2 + \frac{12 du^2}{1 +  3 r^2 } + \frac{36 r^2}{1 +  3 r^2 } {\left( du - \left(2 \xi + 6 r^2 \xi \right) d \varphi\right)}^2\,.
\end{equation}
Here, the last term is positive.
We project in such a way that the auxiliary
metric $\gamma_{\mu \nu}$ reads
\begin{equation}
	\frac{\gamma_{\mu \nu} dx^\mu dx^\nu}{\ell^2}= - dr^2 + \frac{12 du^2}{1 + 3 r^2}\,.
\end{equation}
Here, $r$ runs from $(0, \infty)$ and $u \in (- \infty, \infty)$. Performing the subsequent coordinate transformations 
\begin{equation}
		V =\arctan \left(x+u \right)\,, \quad U =\arctan \left( u-x \right)\,,
\end{equation}
with 
\begin{equation}
	x = \int \frac{\sqrt{1+3r^2}}{\sqrt{12 }} dr
\end{equation}
and
\begin{equation}
	V = T + X\,, \quad U = T- X\,,
\end{equation}
we obtain
\begin{equation}
	\gamma_{\mu \nu} dx^\mu dx^\nu = \frac{\ell^2}{\Omega^2} \left(- dX^2 + dT^2 \right)
\end{equation}
with 
\begin{equation}
	\Omega^2 = \frac{ (1+3r^2)}{12 (1 + {(u-x(r))}^2)(1+{(u+x(r))}^2)}
\end{equation}
As $x \in (0, \infty)$ it follows that $V \geq U$ which in turn implies $X \geq 0$. This leads to the projection diagram \ref{figxiomega0} for the $\xi+\omega = 0$ warped flat quotient.
The conformal factor $\Omega^2  $ goes to $0$ for $r \to \infty$ or $u \to \pm \infty$ and is regular at $r = 0$.
In analogy to the other cases it can be shown that the three-dimensional $\xi+\omega = 0$ warped flat quotient is not a black hole.

\section{Warped Flat Spacetimes in TMG}
\label{TMGfirstorder}

In this section we will determine the asymptotic symmetries of a consistent phase space including the warped flat quotients. To do so, we will particularize to a specific gravity theory admitting them as solutions.
The spacetimes \eqref{wfbhxi} solve the equations of motion of Topologically Massive
Gravity (TMG) \cite{Deser:1981wh} 
\begin{equation}\label{TMGeom}
G_{\mu\nu} + \Lambda g_{\mu\nu}  + \frac{1}{\mu} C_{\mu\nu} = 0 \,,
\end{equation}
with $\mu = \frac{3\sqrt{3}}{\ell}$ and $\Lambda = \frac{1}{\ell^2}$
\footnote{The sign of $\mu$ depends on the orientation of the epsilon-tensor, which 
we have chosen to be $\epsilon^{u \rho \varphi} = - 1/\sqrt{- \mathrm{det}g }$
}. We will find it convenient to work with the first-order formulation of TMG \cite{Carlip:2008qh}, so that we can use methods developed for general Chern-Simons-like theories of gravity in \cite{Bergshoeff:2014bia,Grumiller:2017otl}. An additional reason for working in the first-order formulation is that it is straightforward to obtain a well-defined variational principle, making it possible to single out a particular symplectic structure and thus fixing the ambiguity of adding boundary terms to the action. In fact, the charges as defined in \cite{Bergshoeff:2014bia,Grumiller:2017otl} are already compatible with a well-posed variational principle.

\subsection{First-order Formulation of TMG}
The first-order formulation of TMG is given by the action of the form
\begin{equation}\label{STMG}
S_{\textsc{tmg}} = \frac{1}{8\pi G} \int  \Bigg[ - e_a \wedge R^a + \frac{1}{\mu} f_a \wedge D e^a + \frac{1}{2\mu} \omega_a \left( d \omega^a + \frac13 \epsilon^{abc} \omega_b \wedge \omega_c \right) + \frac{\Lambda}{6} \epsilon_{abc} e^a \wedge e^b \wedge e^c \Bigg],
\end{equation}
where $e^a$ is the dreibein, $\omega^a$ is the dualized spin-connection and $f^a$ is an auxiliary field which enforces the vanishing torsion constraint. The field equations read
\begin{subequations}
	\label{fieldeq}
	\begin{align}
	&\dd e^a + \epsilon^a{}_{bc}\omega^b \wedge e^c = T(\omega)^a = 0 \label{fieldeq1}\\
	&\dd \omega^a + \frac{1}{2} \epsilon^a{}_{bc}\omega^b \wedge \omega^c = R(\omega)^a = -  \epsilon^a{}_{bc}e^b \wedge f^c\label{fieldeq2}\\
	& \frac{1}{\mu} D f^a  + \epsilon^a{}_{bc}e^b \wedge f^c + \frac{\Lambda}{2}  \epsilon^a{}_{bc}e^b \wedge e^c= 0 \label{fieldeq3}\,.
	\end{align}
\end{subequations}
All of the fields appearing in the action \eqref{STMG} are Lorentz-vector valued one-forms, hence we can collect them into a single Chern-Simons-like one-form field with an additional (flavor) index $a^\ione{}^a = \{e^a, \omega^a, f^a\}$. The action \eqref{STMG} is then economically expressed as
\begin{align}\label{CSlikeaction}
S &= \frac{1}{8\pi G} \int \left(
\frac12 g_{\ione\itwo} a^\ione{}^a \wedge \dd a^\itwo{}_a + \frac{1}{6} f_{\ione\itwo\ithr} \epsilon^{abc} a^\ione_a \wedge a^\itwo_b \wedge a^\ithr_c \right) \, .
\end{align}
This is known as the Chern-Simons-like (CS-like) action \cite{Bergshoeff:2014bia,Merbis:2014vja} and can be used to derive for instance asymptotic charges \cite{Bergshoeff:2019rdb} and asymptotic symmetries \cite{Grumiller:2017otl} for a variety of 3D massive gravity models at the same time. Upon choosing the completely symmetric flavor space metric $g_{\ione \itwo}$ and structure constants $f_{\ione\itwo \ithr}$ to be
\begin{subequations}
	\begin{align}
	&g_{e \omega} = -1\,, \quad g_{\omega \omega} = \frac{1}{\mu}\,, \quad g_{ef} = \frac{1}{\mu}\\
	& f_{e \omega \omega} = -1\,, \quad f_{eee} = \Lambda\,, \quad f_{e \omega f} = \frac{1}{\mu}\,, \quad f_{\omega \omega \omega} = \frac{1}{\mu}\,,
	\end{align}
\end{subequations}
one can easily recover \eqref{STMG} from \eqref{CSlikeaction}.

In \cite{Grumiller:2017otl} the formalism to compute charges for asymptotic Killing vectors in CS-like theories was developed. One first specifies certain boundary conditions on the set of fields $a^{\ione}$, consistent with the equations of motion (at least, asymptotically to the relevant order in $r$). The boundary condition contains, besides a specification of what is fixed as one goes towards the boundary, the specification of what is allowed to vary on the boundary and constitutes the state-dependent information of the theory. Then one considers the gauge-like transformations
\begin{equation}
\label{eq:wbh5}
\delta_{\chi} a^\ione = \dd \chi^\ione + f^\ione{}_{\itwo\ithr}\,[a^\itwo,\,\chi^\ithr]\,,
\end{equation}
which leave the boundary conditions invariant, up to transformations of state-dependent functions. In general, not all $\chi^{\ione}$ generate gauge symmetries, some of them are related to second class constraints and hence fixed in terms of $a^\ione$. This is why we referred to them as gauge-like parameters. The gauge-like parameters corresponding to diffeomorphisms take the form
\begin{equation}\label{xiKilvec}
\chi^\ione = a_{\mu}^\ione \zeta^\mu\,.
\end{equation}
One then uses the obtained $\chi^{\ione}$ to compute the asymptotic charges \cite{Bergshoeff:2014bia,Grumiller:2017otl} (on the boundary of the disk at a constant time slice)
\begin{equation}
\label{varboundary}
\delta Q[\chi^\ione] = - \frac{1}{8\pi G} \int_0^{2 \pi} \dd \varphi \; \big(g_{\ione\itwo}\, \chi^\ione\, \cdot \delta a_\vp^\itwo\big)\,.
\end{equation}
Here the dot denotes contraction with the $SO(2,1)$ invariant metric $\eta_{ab}$. This expression should be integrable and finite in order for the boundary conditions to be consistent. The asymptotic symmetry algebra is then spanned by the Dirac brackets of the charges, which is most easily computed as $\{Q[\chi^\ione], Q[\eta^{\itwo}]\} = - \delta_{\eta^{\itwo}} Q[\chi^{\ione}] $.

\subsection{Phase Space}
While the considerations in section \ref{sec:geometry} have been coordinate independent, the computation of asymptotic charges relies on the introduction of a coordinate system ($u, r, \varphi$) and a specification of the fall-off conditions on the metric as $r \to \infty$.
In order to compute the asymptotic symmetry algebra of the warped flat quotient \eqref{wfbhxi}, we find it convenient 
to make a coordinate transformation to a radial variable $r$ defined as
\begin{align}\label{rredef}
\rho = - 6 r^2 \xi \, .
\end{align}
Note that we had assumed $\xi > 0$ without loss of generality and hence this coordinate transformation keeps us in the future wedge with $\rho <0$, where there are never any closed timelike curves.
We obtain the metric
\begin{align}
\label{wfbhmj}
ds^2 = \ell ^2 \left( (12  \xi r )^2 d\varphi^2+ 12 \left(d\varphi  \left(\xi - 6 r^2 \xi +\omega
\right)+du\right)^2 -  dr^2\right)
\,.
\end{align}
The advantage of using this metric over \eqref{wfbhxi} is that now $\xi$ and $\omega$ can be promoted to arbitrary functions of $\varphi$, while still solving the TMG field equations \eqref{TMGeom} (now with $\mu = - \frac{3 \sqrt{3}}{\ell} $, as the coordinate transformation \eqref{rredef} changes the spacetime orientation).\footnote{Here we have taken the orientation of the epsilon-tensor
to be $\epsilon^{u r \varphi} = - \frac{1}{\sqrt{-\det g}}$}

The state-dependent data in this metric are the two functions $\xi$ and $\omega$, which we will now take to be arbitrary functions of $\vp$. The next step consists in finding the Chern-Simons-like fields $a^\ione$ consistent with the TMG field equations and leading to the metric \eqref{wfbhmj} through $g_{\mu \nu} = e_\mu^a e_\nu^b \eta_{ab}$. We choose the following Lorentz frame for the dreibein:
\begin{subequations}
	\label{eq:triadWF}
	\begin{align}
	e_u & = 2 \sqrt{3} \ell\, T_1\,, \\
	e_r & = \ell\, T_0\,, \\
	e_{\vp} & =  2\sqrt{3}\ell \left( (1- 6  r^2)  \xi (\varphi )  +  \omega(\varphi) \right) T_1 - 12 r \ell \xi(\varphi) \, T_2\,.
	\end{align}
\end{subequations}
Here $T_a$ are $SO(2,1)$ generators.
Solving \eqref{fieldeq1} we find for the components of the spin-connection
\begin{subequations}
	\label{eq:spinconWF}
	\begin{align}
	\omega_u & = - 6 \, T_1\,, \\
	\omega_r & =  \sqrt{3} \, T_0\,, \\
	\omega_{\vp} & =  6 \left((1+6 r^2)\xi (\varphi ) - \omega (\varphi)\right)\, T_1 - 12 \sqrt{3} r \xi (\varphi )\, T_2 \,.
	\end{align}
\end{subequations}
Equation \eqref{fieldeq2} can be solved to find the components of the auxiliary field as
\begin{subequations}
	\label{eq:auxfieldWF}
	\begin{align}
	f_u & =\frac{15 \sqrt{3}}{\ell} \, T_1\,, \\
	f_r & = - \frac{9}{2 \ell}\, T_0\,, \\
	f_{\vp} & = - \frac{15 \sqrt{3}}{\ell} \left( (6  r^2 -1) \xi(\varphi) - \omega (\varphi )\right) \, T_1 + \frac{54 r \xi(\varphi)}{\ell}\, T_2 \,.
	\end{align}
\end{subequations}
This we will consider to be our boundary conditions for the Chern-Simons-like fields $a^{\ione}$. This terminology might be confusing, as these are exact solutions to the TMG field equations and not defined in terms of an asymptotic expansion close to the boundary at $r \to \infty$. It is, however, a common feature in three-dimensional gravity that the asymptotic expansions are finite. For instance, the Fefferman-Graham expansion in AdS$_3$ terminates at the second order.

\subsection{Asymptotic symmetry transformations}\label{sec:AST}
We will now consider gauge-like transformations \eqref{eq:wbh5} which leave $a^{\ione}$ invariant up to transforming the state-dependent functions $\xi$ and $\omega$. These parameters will correspond to asymptotic symmetry transformations with asymptotic Killing vectors obtainable through \eqref{xiKilvec}. By explicitly solving \eqref{eq:wbh5}, we find that they are given in terms of two arbitrary functions of $\vp$, denoted here by $T(\vp)$ and $Y(\vp)$, as:
\begin{subequations}\label{xi}
	\begin{align}
	& \chi^e = \ell f(\vp) T_0
	+ \left( - 12 \ell r Y(\vp) \xi(\vp) + \frac{\ell f'(\varphi )}{12 \xi (\varphi )} \right)  T_2 \\
	&\qquad +\left(- 12 \sqrt{3} r^2 \ell  \xi(\varphi ) Y(\varphi ) + 2 \sqrt{3} \ell \left( T(\varphi ) +  (\xi (\varphi ) + \omega(\vp)) Y(\varphi )\right) + \frac{ \ell r f'(\varphi)}{2\sqrt{3} \xi(\vp)} 
	\right) T_1\,, \nonumber
	\\
	& \chi^\omega = \sqrt{3} f(\varphi ) T_0
	+ \frac{-144 r Y(\vp) \xi(\vp)^2 + f'(\varphi )}{4 \sqrt{3} \xi (\varphi )} T_2 \\
	& \qquad + \left(- 6 T(\varphi ) + 6 Y(\vp) \left( (1 + 6 r^2) \xi(\vp) - \omega(\vp) \right)  - \frac{r f'(\vp) }{2 \xi(\vp)} 
	\right) T_1\,, \nonumber
	\\
	& \chi^f = -\frac{9 f(\varphi)}{2 \ell} T_1
	+\left(\frac{54 r Y(\vp) \xi(\vp) }{\ell} -  \frac{3 f'(\varphi )}{8 \ell \xi (\varphi )} \right) T_2 \\
	&\qquad+ \frac{5 \sqrt{3}}{4 \ell \xi(\vp)} \left( 12\xi(\vp) (T(\vp) + Y(\vp) ((1-6 r^2)\xi(\vp) + \omega(\vp) ) + r f'(\vp) \right) T_1\,. \nonumber
	\end{align}
\end{subequations}
where $f(\vp)$ solves the differential equation
\begin{equation}
f''(\vp) - \frac{ \xi'(\vp)}{\xi(\vp)} f'(\vp) - 144 \xi(\vp)^2 f(\vp) = 0\,.
\end{equation}
This solution is given by
\begin{equation}
\label{fsol}
f(\vp) =  \left( c\,  e^{-12 \int_{0}^{\vp} \xi(\vp') \dd \vp' } + d\,  e^{12 \int_{0}^{\vp} \xi(\vp') \dd \vp' } \right)\,.
\end{equation}
for arbitrary constants $c$ and $d$.

These asymptotic gauge transformations preserve \eqref{eq:triadWF}, \eqref{eq:spinconWF} and \eqref{eq:auxfieldWF} provided that
\begin{subequations}
\begin{align}\label{delsol}
	\delta_\chi \xi & = \partial_\vp( \xi(\vp) Y(\vp) ) \,, \\
	\delta_\chi \omega & = T'(\vp) + \partial_\vp (\omega(\vp) Y(\vp))\,.
\end{align}
\end{subequations}
The asymptotic Killing vectors connected with boundary conditions preserving gauge transformations can be calculated via \eqref{xiKilvec}, and read
\begin{align}\label{kilvec}
\zeta(T, Y, c, d) &=  \left(T(\varphi ) + \frac{\left((1 + 6 r^2) \xi + \omega\right)f'(\varphi )}{144 r \xi^2} \right) \partial_u   + \left(Y(\varphi ) - \frac{f'(\varphi )}{144 r   \xi^2} \right)\partial_\vp \\
& \quad +
f(\vp) \partial_r \nonumber \,,
\end{align}
with $f(\varphi)$ given by \eqref{fsol}. Note that \eqref{xiKilvec} holds only for the dreibein and the auxiliary field. The equation for $\omega$ holds up to a term proportional to $\xi^\omega =  \frac{f'(\vp)}{12 r\xi} T_1$. This term can be removed by a local Lorentz transformation.

Before we compute the asymptotic charges and their symmetry algebra, let us look at the exact Killing vectors for the warped flat quotient \eqref{wfbhmj} with constant $\xi$ and $\omega$. 
The 4 Killing vectors are given by:
\begin{align}
\partial_u\,, \qquad \partial_\vp\,, \qquad  e^{\pm 12 \xi \vp} \left( \pm \frac{1}{12 r \xi} (\xi(1+ 6r^2) + \omega) \partial_u \mp \frac{1}{12 r \xi} \partial_\vp + \partial_r \right)\,.
\end{align}
We observe that the first two correspond to linear combinations of $I_0$ and $H$ in \eqref{globalKV} and are globally well-defined after the identifications. 
The last two are only globally well-defined and single-valued in the domain $0 \leq \vp <2\pi$ when 
\begin{equation}\label{xivac}
\xi = \frac{i}{12}\,.
\end{equation}
In analogy with AdS$_3$ gravity, we identify the vacuum as (one of the metrics) with that value of the parameter $\xi$.

\subsection{Charge algebra}
We now compute the charges from the general formula for Chern-Simons-like theories \eqref{varboundary}. Using \eqref{eq:triadWF}, \eqref{eq:spinconWF}, \eqref{eq:auxfieldWF} and \eqref{xi}, the variation of the boundary charges \eqref{varboundary} is integrable and finite, but it only depends on the gauge parameters $T$ and $Y$; the dependence on $f(\vp)$ drops out completely
\begin{equation}
Q[T, Y] =\frac{\sqrt{3} \ell}{\pi G} \int d \varphi \left(2  \xi (\varphi ) T(\varphi ) + Y(\varphi )  \xi (\varphi ) \left( 3  \xi(\varphi) + 2 \omega (\varphi ) \right) \right) \,.
\end{equation}
Using $\delta_\epsilon Q[\eta] = \{Q[\epsilon], Q[\eta]\}$ we compute the Dirac bracket algebra of the Fourier modes of the charges, 
defined as $P_n = Q[T = e^{i n \varphi}, Y = 0]$ and
$L_n = Q[T = 0, Y = e^{i n \varphi}]$. Replacing $i \{\,, \, \} \rightarrow [ \, , \,]$
we find
\begin{align}
\label{wcftvan}
[ L_n, L_m ]&=  (n-m) L_{n+m}\nonumber \\
[L_n, P_m]&=  - m P_{n+m} \\
[P_n, P_m] &=0\,.\nonumber
\end{align}
This is a warped conformal symmetry algebra, but with vanishing central charge and  vanishing $\hat{u}(1)$ level.

The asymptotic symmetry algebra is an infinite-dimensional lift of the charges associated to the global Killing vectors $I_0$ and $H$, whereas the asymptotic Killing vectors associated to the function $f(\vp)$ in \eqref{fsol} do not have any corresponding asymptotic charges and neither appear in the transformation rules of $\xi$ and $\omega$. Hence these asymptotic Killing vectors do not play any role in the asymptotic symmetry algebra.

\subsection{Thermodynamics}
In this section we will discuss the thermodynamics associated to the horizon of the warped flat quotient. We have seen in section \eqref{causalwf} that for $\xi + \omega = 0$ the spacetime has to be cut off at $\rho = 0$ due to the occurence of closed timelike curves. Hence, in this case the spacetime does not possess a horizon at $\rho = 0$ anymore. Naturally,
 we will thus  restrict our considerations in this section to the case $\xi + \omega >0$.
For constant $\xi(\vp) = \xi$ and $\omega(\vp) = \omega$ the mass and 
angular momentum read
\begin{subequations}
	\label{L0P0}
	\begin{align} 
	M &= P_0 = \frac{4 \sqrt{3} \ell \xi}{G} \\
	J &= L_0 = \frac{2 \sqrt{3} \ell \xi ( 3 \xi + 2 \omega)}{G}\,,
	\end{align}\end{subequations}
The mass $M$ and the angular momentum $L_0$ are positive and bounded from below due to our assumptions $\xi>0$ and $\xi+\omega >0$. 

The angular velocity and the temperature read
\begin{align}
\label{tempang2}
\Omega_H &= - \frac{g_{t \varphi}}{g_{\vp \vp}} \bigg|_{r = 0} = -\frac{1}{\xi + \omega}\\
T_H &= \frac{\kappa}{2 \pi}  =  \frac{6}{\pi}\frac{ \xi}{\xi+\omega} \,,
\end{align}
The entropy is given as \cite{Kraus:2005zm,Tachikawa:2006sz,Bouchareb:2007yx,Solodukhin:2005ah,Detournay:2012ug}
\be 
S = S_E +  S_C
\,,
\ee
where \footnote{The sign of the Chern-Simons contribution $S_C$ depends on
orientation.}
\begin{align}
S_E &= \frac{2 \pi R_0}{4 G} \bigg |_{r \to 0} = \frac{\sqrt{3} \pi \ell (\xi + \omega)}{G} \\
S_C & =  \bigg \vert \frac{M_0}{N} \bigg \vert \frac{\pi  R_0^2 \partial_r N_\vp}{4 G \mu} \, \bigg |_{r \to 0} = + \frac{\pi \ell (\xi - \omega)}{\sqrt{3} G }  \,.
\end{align}
Here, the functions are defined by the decomposition of the metric in the following form
\be
ds^2=-N^2 du^2 + \frac{dr^2}{M_0^2}
+R_0^2 \left(d\varphi+ N_\varphi du\right)^2
,
\ee
which explicitly reads
\begin{subequations}
\begin{align}\label{M0N}
N^2 &= - \frac{144 r^2 \ell^2 \xi^2}{(1+36 r^4)\xi^2 +2 (1-6 r^2)\xi \omega +\omega^2} \,, \\
M_0^2 &= - \frac{1}{\ell^2} \,, \\
R_0^2 &= 12 \ell^2 ((1+36 r^4) \xi^2 +2(1-6 r^2)\xi \omega +\omega^2) \,, \\
N_\varphi &= \frac{\xi+\omega -6 r^2 \xi}{(1+36 r^4)\xi^2 + 2 (1-6 r^2)\xi \omega +\omega^2}\,.
\end{align}
\end{subequations}
We obtain the entropy
\be
\label{TMGentropy2}
S=\frac{2 \pi \ell  (2\xi +\omega )}{\sqrt{3} G}\,,
\ee 
which then satisfies a first law of the following form
\begin{equation}
\delta S = - \beta \delta M - \beta \Omega \delta J\,.
\end{equation}
The sign of the first law is reminiscent of the first law at the inner horizon of black holes \cite{Detournay:2012ug} or for cosmological horizons \cite{Bagchi:2012xr}.
We see that the entropy \eqref{TMGentropy2} is manifestly positive as $\xi+\omega >0$ and $\xi >0$.
When expressed in terms of the mass $P_0$ and angular momentum $L_0$ the entropy reads
\begin{equation}\label{WFS}
S = \left( \frac{\pi }{12} P_0 + \frac{2\pi \ell}{\sqrt{3} G } \frac{L_0}{P_0} \right)
\end{equation}
The goal in the next section will be to reproduce \eqref{WFS} using the warped conformal symmetries \eqref{wcftvan} with vanishing central charges.

\section{Warped Conformal Field Theories at vanishing level}\label{sec:WCFT}
The algebra \eqref{wcftvan} describes a warped conformal field theory with
no central extensions. In this section we reproduce \eqref{WFS} by using a Cardy-like entropy formula for warped conformal field theories of the form \eqref{wcftvan}. This formula can be derived by a slight generalization of the derivation provided in \cite{Detournay:2012pc}, see appendix \ref{appdercardy} for the details.
For this we will consider a warped conformal
field theory on the torus in Lorentzian signature described by coordinates $(t, \varphi)$ chosen
such that the symmetries are
\begin{equation}
    \label{symmcyl}
\varphi^\prime = f(\varphi)\,, \quad t^\prime = t - g (\varphi)
\end{equation}
The identifications of the coordinates read
\begin{equation}
(t, \varphi) \sim (t, \varphi + 2 \pi) \sim (t+ i \beta, \varphi + \theta)\,.
\end{equation}
These transformations \eqref{symmcyl} are generated by infinitesimal vector fields
$l_n = i e^{i n \varphi} \partial_\varphi$ and $p_n = e^{i n \varphi} \partial_t$ fulfilling
\begin{align}
    \label{a2eq00}
         [ l_m, l_n ] &=  (m-n) l_{m+n} \nonumber \\
         [ l_m, p_n ] &= - n p_{m+n}   \\
        [ p_m, p_n ] &= 0\,. \nonumber
    \end{align}
    The algebra of the charges $L_n, P_n$ on the torus is given by \eqref{a2eq00}
    up to central extensions.
The algebra \eqref{a2eq00} admits three non-trivial central extensions
$(c, \kappa, k)$
    \begin{align}
    \label{WCFT}
         [ L_m, L_n ] &=  (m-n) L_{m+n} + \frac{c}{12} (m^3-m) \delta_{n+m, 0} \,, \nonumber \\
         [ L_m, P_n ] &= - n P_{m+n} - i \kappa (m^2 +m) \delta_{m+n, 0} \,,  \\
        [ P_m, P_n ] &= \frac{k}{2} m \delta_{n+m, 0}\,.\nonumber
    \end{align}
The partition function at inverse temperature $\beta$ and angular potential
$\theta$ reads
\begin{align}
    \label{eq14}
        Z(\beta, \theta) &= \mathrm{Tr}\exp\left(-\beta P_0 + i \theta L_0 \right) \,.
     \end{align}
Under a modular transformation the partition function transforms as
\begin{equation}
Z(\beta, \theta)= \exp \left( \frac{i \beta^2}{4 \theta} k  \right) Z \left(\frac{2 \pi \beta}{\theta}, - \frac{4 \pi^2}{\theta} \right)\,.\nonumber
\end{equation}
The modular properties of this partition function
were discussed in \cite{Detournay:2012pc}
for warped conformal field theories with $(c, 0, k)$. For $L_0$ bounded from below
we find in the regime of small imaginary $\theta$ (provided that
the vacuum is gapped) that
    \begin{equation}
        \label{eq15}
            Z(\beta, \theta) \approx \exp \left(\frac{i \beta^2}{4 \theta} k  \right)\exp\left(- \frac{2 \pi \beta}{\theta} P_0^{\mathrm{vac}} - i \frac{4 \pi^2}{\theta} L_0^{\mathrm{vac}} \right)\,.
        \end{equation}
Using the thermodynamic relation $S = (1 - \beta \partial_\beta - \theta \partial_\theta) \ln(Z)$
we find on the one hand for WCFTs with $(c, \kappa, k)$,
provided that $k \neq 0$, that
\begin{align}
    \label{a2eq210}
    S = - \frac{4 i P_0 P_0^{\mathrm{vac}} \pi}{k} + 4 \pi \sqrt{- \left(  L_0 - \frac{P_0^2}{k} \right) \left(  L_0^\mathrm{vac} - \frac{{P_0^\mathrm{vac}}^2}{k}
    \right)
    }\,,
    \end{align}
while on the other hand for warped conformal field theories with generic $(c, \kappa, 0)$
\begin{equation}
    \label{a2eq190}
    S = - 2 \pi i \left( \frac{L_0^\mathrm{vac}}{P_0^{\rm vac}} P_0 + P_0^\mathrm{vac} \frac{L_0}{P_0} \right) \,,
\end{equation}
as advertised in \eqref{WFEntropy}. Thus, in our case of $(c, \kappa, k)$  being $(0,0,0)$, we obtain \eqref{a2eq190}.

At this point, we need to identify the vacuum charges. We saw in section \ref{sec:AST} that the family of metrics \eqref{wfbhmj} generically exhibits two globally defined Killing vectors, except at the special value \eqref{xivac}
\begin{equation}
\xi^{\mathrm{vac}} = \frac{i}{12}
\end{equation}
where a symmetry enhancement occurs and which we thereby identify with the vacuum value. Using \eqref{L0P0} this leads to
\begin{equation}\label{P0vac}
	P_0^\mathrm{vac} = \frac{ i \ell}{\sqrt{3} G}\, .
\end{equation}
However, this argument alone is not sufficient to specify the vacuum value of $\omega$. We will now argue, in four independent ways, that the vacuum value of $\omega$ is given by
\begin{equation}\label{omegavac}
\omega^{\rm vac} = - \frac{i}{12} \,,
\end{equation}
leading to 
\be\label{L0vac}
L_0^{\rm vac} = - \frac{\ell}{24 \sqrt{3} G}\,.
\ee
First, for $\xi = \frac{i}{12}$ and  $\xi + \omega = 0$, the coordinate transformation \eqref{coordinatetraf} does not imply any identification and hence does not quotient the spacetime. Secondly, as discussed in section 2, the $\xi + \omega = 0$ solution does not possess a horizon because the spacetime is truncated at $\rho =0$ to excise closed time-like curves. The solution thus has no temperature and vanishing entropy, which is another indication that this solution corresponds to the ground state. 

Thirdly, as shown in appendix \ref{wfqe}, the warped flat quotient can be viewed as a deformation of a 
locally flat metric \eqref{ds2flatMJ} depending on two parameters ($M$ and $J_{\rm wf}$)
 and $1/\ell$ playing the role of a deformation parameter. 
 A natural guess for the ground state is then to take the vacuum metric as the one corresponding to a deformation of global Minkowski space, which is obtained from \eqref{ds2flatMJ} by choosing $M = -1$ and $J_{\mathrm{wf}} = 0$.
 This corresponds exactly to $\xi^\mathrm{vac} = - \omega^\mathrm{vac} = i/12$.
%The values of $M$ and $J_{\rm wf}$ for which the undeformed metric becomes Minkowski spacetime correspond exactly to $\xi^{\rm vac} = - \omega^{\rm vac} = i/12$.
 A similar argument to fix $L_0^{\rm vac}$ geometrically in the case of warped AdS$_3$ was put forward in section 5.5 of \cite{Detournay:2012pc}.

Finally, we can derive \eqref{L0vac} from a limit of a warped CFT with non-vanishing U(1) level. In that case, the vacuum charges are given by \cite{Detournay:2012pc}
\be\label{vacrel}
L_0^{\rm vac} = -\frac{c}{24} + \frac{q^2}{k} \,.
\ee
See appendix \ref{appdercardy} for a review of this argument. 
For a WCFT dual to WAdS$_3$ spacetimes in TMG, the values of $q$, $c$ and $k$ are known. We have repeated the asymptotic symmetry analysis for these spacetimes with positive cosmological constant in appendix \ref{app:WAdS}, of which the warped flat solutions can be obtained by the limit $\nu^2 \to 3$. The result is 
\be\label{vacL0}
L_0^{\rm vac} = - \frac{\ell}{24 \nu G}\,.
\ee
%We will assume that this relation holds for holographic WCFTs dual to TMG, i.e. a gravity theory where the diffeomorphism anomaly is captured by a gravitational Chern-Simons term with coupling constant $\nu$. 
From this result, we see that there is a smooth limit of \eqref{vacrel} as $\nu^2 \rightarrow 3$ (even though both $c$ and $1/k$ blow up in that limit) so that at $\nu^2 = 3$ we recover \eqref{L0vac}.\footnote{The sign of $\nu$ is related to the orientation and is fixed to $\nu = \sqrt{3}$ in our case.} 
%\be\label{L0vac}
%L_0^{\rm vac} = - \frac{\ell}{24 \sqrt{3} G}\,.
%\ee

Plugging \eqref{L0vac} and \eqref{P0vac} in \eqref{a2eq190}, we reproduce exactly the entropy \eqref{WFS}
\begin{equation}
	S = \left( \frac{\pi }{12} P_0 + \frac{2\pi \ell}{\sqrt{3} G } \frac{L_0}{P_0} \right) \,.
	\end{equation}

\section{Discussion}\label{sec:discussion}

We have found a consistent set of boundary conditions accommodating the family of solutions (\ref{wfbhxi}). Their ASA was shown to consist in a centerless WCFT algebra (\ref{wcftvan}). The corresponding field theory has a degeneracy of states given by (\ref{a2eq190}) which matches the Bekenstein-Hawking entropy (\ref{WFS}) of the corresponding solutions.

\noindent We conclude with some comments:

\begin{enumerate}
\item The gravitational theory we have considered is TMG, which we have expressed in first-order formulation. This allowed us to use an unambiguous definition of the charges inherent to Chern-Simons-like theories. It would be interesting to reproduce our results using Iyer-Wald \cite{Iyer:1994ys} or Barnich-Brandt \cite{Barnich:2001jy} charges for TMG in metric formulation. Boundary terms needed for a well-defined variational principle for our boundary conditions might be required to make the asymptotic charges well-defined\footnote{A similar observation is probably valid for more general boundary conditions in AdS$_3$ \cite{Grumiller:2016pqb}}. Extensions to other higher-curvature theories/theories with matter could also be performed.

\item The entropy formula (\ref{wcftvan}) we derived for a centerless WCFT shares interesting similarities with that of a BMS$_3$ algebra. 
% Indeed, let us perform the (non-local) change of coordinate $u = P_0 \tilde{u}$ and let $\tilde{P_0} := Q_{\partial_{\tilde{u}}}$.
% It follows that $\delta \tilde{P_0}  = P_0 \delta P_0$, and hence $\tilde{P_0} \sim P_0^2$. This means that in terms of this ``quadratic charge", the entropy reads as 
% $S = \alpha_1 \sqrt{\tilde{P_0}} +  \alpha_2 \frac{L_0}{\sqrt{\tilde{P_0}}}$. 
Indeed, let's consider the warped flat quotients in the coordinates of section \ref{wfqe}. 
In particular, defining $\tilde{P_0} = Q_{\partial_t} $, we find that $\tilde{P_0} = \frac{ \sqrt{3} G}{2} P_0^2$. In terms of these charges, the entropy reads (for $\tilde P_0^{\rm vac} < 0$ and $\tilde P_0 >0$)
\begin{equation}
S = 2\pi \left( L_0 \sqrt{ - \frac{  \tilde P_0^{\rm vac}}{\tilde P_0} }  -  L_0^{\rm vac}  \sqrt{ - \frac{ \tilde P_0}{\tilde P_0^{\rm vac}} }   \right) \,.
\end{equation}

This is precisely the BMS-Cardy entropy formula (see third column of Table \ref{cardytable}) after having used the relationship between vacuum charges and central extensions in a BMS$_3$-invariant field theory (see Sect 3.1 of \cite{Jiang:2017ecm}).
%This has the same form as the BMS-Cardy entropy formula (see third column of Table \ref{cardytable}). 
This is reminiscent of the quadratic ensemble in WCFTs, in which the degeneracy of states takes the form of a Cardy formula even in the absence of conformal symmetry \cite{Detournay:2012pc}. In this case, it is not clear whether a BMS symmetry is present or not. It would be interesting to perform an asymptotic symmetry analysis in this ensemble.
\item 

The matching of the field theory and gravitational entropies required to identify the vacuum charges. We gave several arguments for these values. However, from a purely field-theoretical point of view, these values should have been vanishing, because of the vanishing of all central extensions (see \ref{a2eq26})%\footnote{To be more precise, the vacuum values \eqref{P0vac} and \eqref{L0vac} are obtained from \eqref{a2eq26} by taking $\delta \to \infty$ as $k \to 0$ such that $\delta k = q$ remains finite. In addition, one should take $c\to \infty $ such that $\delta^2 k - \frac{c}{24}$ remains finite.}
. The determination of the vacuum charges on the cylinder using a warped conformal transformation assumes a state-operator correspondence that might not be true in this case. This adds to the fact that highest-weight representations are generically non-unitary for WCFTs with vanishing $k$, which does not exclude the existence of unitary induced representations. Understanding these aspects of WCFTs is certainly worth exploring.

\item WF spacetime has a 4-parameter family of isometries. When quotienting out the global spacetime by a one-parameter discrete subgroup of the isometry group generated by a Killing vector $\Xi$, the resulting geometry will generically have as Killing vectors all original Killing vectors commuting with $\Xi$. This means that, in order to get a geometry with two Killing vectors, one could quotient out WF by a vector of the form $\alpha X + \beta I_0$ where $X$ is any Killing vector. The spacetimes studied here correspond to %$\beta = -\frac{\xi + \omega}{2}$ and 
$X = H$ and non-vanishing $\alpha$ and $\beta$, while self-dual warped flat is obtained for $\alpha=0$ \cite{Anninos:2009yc}.
It might be interesting to investigate whether more general quotients could lead to physically meaningful solutions.

\end{enumerate}

\section*{Acknowledgements}
We thank Hamid Afshar, Dionysios Anninos, Geoffrey Comp\`ere, Daniel Grumiller, Rodrigo Olea, Mateusz Piorkowski and Georg Stettinger for discussions. We would especially like to thank Gerard Clement for discussions on the properties of warped flat spacetimes. RW especially thanks Piotr Chru\'{s}ciel for interesting and helpful discussions about projection diagrams and Friedrich Sch\"{o}ller for countless helpful discussions. We would also like to thank Tatsuo Azeyanagi and Stefan Prohazka for collaborations at an earlier stage of the project.

SD is a Research Associate of the Fonds de la Recherche Scientifique F.R.S.-FNRS (Belgium). SD was supported in part by the ARC grant ``Holography, Gauge Theories and Quantum Gravity Building models of quantum black holes'', by IISN -- Belgium (convention 4.4503.15) and benefited from the support of the Solvay Family. SD acknowledges support of the Fonds de la Recherche Scientifique F.R.S.-FNRS (Belgium) through the CDR project C 60/5 - CDR/OL. WM is supported by the ERC Advanced Grant \emph{High-Spin-Grav} and by FNRS-Belgium (convention FRFC PDR T.1025.14 and convention IISN 4.4503.15). GN is supported by Simons Foundation Grant to HMI under the program ``Targeted Grants to Institutes''. RW is supported in part by a DOC Fellowship of the Austrian Academy of Sciences, by the Austrian Science Fund FWF under the Doctoral Program W1252-N27– 15 –
Particles and Interactions and the Christiana H\"{o}rbiger stipend.

Part of this work has been carried out during the Higher Spins and Holography workshop at the Erwin Schr\"{o}dinger Institute (March 11 - April 5 2019), which we would like to acknowledge for support.

\appendix
\section{Limits from WAdS or WdS to Warped Flat Spacetimes}
\label{app:WFaslimits}
\subsection{Limits from WAdS or WdS}
We can either see the warped flat metric \eqref{eqmetricvacuum} as a limit of global WAdS or WdS 
(both with positive cosmological constant).
Following \cite{Anninos:2009jt}, approaching $\nu^2\rightarrow 3$ from $\nu^2>3$,
we start with the global WAdS metric with positive cosmological constant
  \be
  \label{WAdS}
  ds^2 = \frac{\ell^2}{\nu^2-3}\left[-\left(1+r^2\right)d\tau^2 + \frac{dr^2}{1+r^2} + \frac{4\nu^2}{\nu^2-3}\left(du +r d\tau\right)^2\right]\,,
  \ee 
 send $r\rightarrow \sqrt{\nu^2-3}x,~\tau \rightarrow \sqrt{\nu^2-3} \tau,~u \rightarrow (\nu^2-3)y$ and then take the $\nu^2\rightarrow 3$ limit to get the metric in equation (\ref{eqmetricvacuum}). Alternatively, for $\nu^2<3$ we start with WdS in static coordinates
 \be
 ds^2 = \frac{\ell^2}{3-\nu^2}\left[-\left(1-r^2\right)d\tau^2 + \frac{dr^2}{1-r^2} + \frac{4\nu^2}{3-\nu^2}\left(du +r d\tau\right)^2\right]
 \ee and repeat the same limit but replacing $(\nu^2-3)$ with $(3-\nu^2)$ in the transformations to get metric in equation (\ref{eqmetricvacuum})

Next, we discuss how the isometries \eqref{isom} are related to the isometries of global WAdS$_3$ with positive cosmological constant. The Killing
 vectors of global WAdS$_3$ \eqref{WAdS} read \cite{Anninos:2009jt}:

\bea
\tilde{J}_1 &=& 2\frac{r}{\sqrt{1+r^2}}\sin{\tau}  \partial_\tau -2 \sqrt{1+r^2} \cos{\tau}\partial_r + \frac{2 \sin{\tau}}{\sqrt{1+r^2}} \partial_u \nonumber\\
\tilde{J}_2 &=& -2\frac{r}{\sqrt{1+r^2}}\cos{\tau}  \partial_\tau -2
\sqrt{1+r^2}\sin{\tau} \partial_r - \frac{2 \cos{\tau}}{\sqrt{1+r^2}} \partial_u \nonumber\\
\tilde{J}_0&=& 2 \partial_\tau \nonumber\\
J_2&=& 2 \partial_u.
\eea
The $\tilde{J}_i$'s form the $SL(2,R)_L$ and $J_2$ is the $U(1)$. The appropriate limit (as suggested in \cite{Anninos:2009jt}) involves sending
$\tau \rightarrow \tau \epsilon, u \rightarrow y\epsilon^2,
r\rightarrow x \epsilon$, where $\epsilon=\sqrt{3-\nu^2}$ giving
\bea\label{contract}
-\half \epsilon \tilde{J}_1 &=&  \partial_x -  \tau \partial_y =
\half(a_- - a_+ )
\nonumber\\
-\half \left(\tilde{J}_2 +J_2\right) &=&  \tau \partial_x + x \partial_\tau - \half \left(\tau^2+x^2\right)\partial_y =H\nonumber\\
\frac{\epsilon^2}{2} \left(\tilde{J}_2-J_2\right) &=&  -2\partial_y
=I_0
 \nonumber\\
\half \epsilon \tilde{J}_0 &=& \partial_\tau=\half(a_- + a_+ )\,,
\eea thus reproducing the $P_2^c$ algebra \eqref{isomalgebra}.

\subsection{Limit from WAdS black holes with positive cosmological constant}
\label{app:BHaslimits}
Following \cite{Compere:2009zj}, we consider the warped black hole solutions of TMG \emph{at positive cosmological constant}\footnote{This metric is obtainable from equation (1.2) in \cite{Compere:2009zj} upon analytically continuing $\nu \to i \nu$ and $\ell \to i \ell$.}
\be\label{compdet}
ds^2 = dt^2 - 4 \frac{\nu}{\ell} r dt d\theta
+\frac{dr^2}{\frac{\nu^2-3}{\ell^2}r^2-12 m r + \frac{4 j \ell}{\nu}}
+\left[
\frac{3}{\ell^2}\left(\nu^2+1\right) r^2
+12 m r -\frac{ 4 j \ell}{\nu}
\right]d\theta^2 \,,
\ee 
which solve the TMG equations of motion\footnote{Here we have chosen the orientation for the epsilon tensor to be $\epsilon^{t r \theta} = \frac{1}{\sqrt{-\mathrm{det}g}}$.}
\be \label{TMGeom2}
G^\mu_\nu+\Lambda \delta^\mu_\nu + \frac{1}{\mu} C^\mu_\nu=0\,,
\ee 
with $\Lambda=\frac{1}{\ell^2}$ and $\mu= \frac{3\nu}{\ell}$. Here, $(t, r, \theta) \sim
(t, r, \theta + 2 \pi)$.
If we send $r \rightarrow - r, m \rightarrow - m,  \theta \rightarrow - \theta$ the metric, the identifications and the orientation stay the same. 
Thus, $(m, j)$ and $(- m, j)$ describe spacetimes which are isometric with the same orientation. 
Hence, in the following we restrict to $m >0$ without loss of generality.
The local isometries are given by
\bea
T_0&=& \partial_t\,,\quad
L_0 = \partial_\theta \,, \nonumber\\
L_{\pm} &=&
\frac{e^{\pm 2\theta   \sqrt{9m^2 -\frac{j(\nu^2-3)}{\nu \ell}}}}
{\sqrt{4j \ell^3 + \nu r\left[-12 \ell^2 m + (\nu^2-3) r \right]}} \nonumber\\
& &
\left\{
\left(2jl- 3 m \nu r\right) \partial_t
\pm \frac{1}{2 \ell \nu} \sqrt{9m^2 -\frac{j(\nu^2-3)}{\nu \ell}}\left[
4j\ell^3 - 12 \ell^2 m \nu r +\nu(\nu^2-3) r^2
\right]\partial_r \right. \nonumber\\
& &\left.
+ \frac{1}{4 \ell}\left[6\ell^2 m+
(3-\nu^2) r
\right] \partial_\theta
\right\}.
\eea With this choice of normalization, all generators are smooth in the limit of $\nu^2 \rightarrow 3 $. Before taking the limit, we note that the above vectors satisfy the commutators
\bea
[L_+,L_-]&=&
- \frac{\nu^2-3}{4\nu \ell^2} \sqrt{9m^2 -\frac{j(\nu^2-3)}{\nu \ell}}L_0
-\frac{3m \sqrt{9m^2 -\frac{j(\nu^2-3)}{\nu \ell}}}{\ell} T_0
\nonumber\\
~[L_0,L_{\pm}]&=& \pm 2\sqrt{9m^2 -\frac{j(\nu^2-3)}{\nu \ell}} L_{\pm}.
\eea Considering the limit  $\nu^2 \rightarrow 3$, we find the commutation relations
\be
[L_+,L_-]= -\frac{9 m^2}{\ell} T_0,\quad
[L_0,L_{\pm}] =\pm 6 m L_{\pm},
\ee where all other commutation relations vanish. Renaming $H = -\frac{L_0}{6 m}$, $I_0 = -\frac{9 m^2 T_0}{\ell}$ and $a_\pm = L_\pm$ we obtain the algebra \eqref{isomalgebra}. For completeness, the expressions for $L_\pm $ for $\nu=\sqrt{3}$ are
\be
L_\pm =
 \frac{e^{\pm 6 m \theta}}{\sqrt{j \ell -3 \sqrt{3} m r}}
\left[\left(j-\frac{3\sqrt{3}m r}{2 \ell}\right) \partial_t
\pm m \left(\sqrt{3} j \ell -  9m r\right) \partial_r
+\frac{3m}{4} \partial_\theta
\right].
\ee

The metric for $\nu=\sqrt{3}$ is
\be
ds^2 = \frac{dr^2}{-12 m r + \frac{4 j \ell}{\sqrt{3}}}+dt^2 -  \frac{4\sqrt{3}}{\ell} r dt d\theta
+\left[
\frac{12}{\ell^2} r^2
+12 m r -\frac{ 4 j \ell}{\sqrt{3}}
\right]d\theta^2.
\ee
The coordinate transformations to the metric \eqref{wfbhxi} in $u, \rho, \varphi$ coordinates is given by
\bea
t=-\ell \sqrt{12} u ,\,\,\,
\theta = - \ell \varphi,\,\,\,
r=-\ell (\rho +\omega+\xi),\,\,\,
j=-\frac{6\sqrt{3} \xi}{\ell}(\xi+\omega),\,\,\,
m=2 \xi/\ell\,.
\eea
The two Killing horizons of \eqref{compdet} are located at the roots of $g^{rr}=0$ 
\be
r_\pm =
\frac{2}{\nu^2-3}
\left[
 3 m \ell^2 \pm \ell^2 \sqrt{9m^2 -\frac{j(\nu^2-3)}{\nu \ell}}
 \right]\,.
\ee  We define $\epsilon = \nu-\sqrt{3}$ and take $\epsilon\rightarrow 0$ limit, leading to
\bea
r_+ =\frac{2\sqrt{3}m \ell^2}{\epsilon}+O(\epsilon^0)\,,
\quad\quad
r_- = \frac{j \ell}{3\sqrt{3}m } +O(\epsilon^1)\,.
\eea
Thus, we see that the warped flat quotient is obtained from the WAdS$_3$ black holes \eqref{compdet} by taking  $r_+$ to infinity with $r_-$ fixed, analogous to how flat space cosmologies are obtained from a limit of the BTZ black holes.

\subsection{Warped flat limit in the quadratic ensemble}\label{wfqe}
The warped AdS$_3$ spacetime of the previous section can equivalently be understood as a deformation of a locally (Euclidean) de Sitter solution, which defines the (de Sitter analogue of the) quadratic ensemble of \cite{Detournay:2012pc}:
\begin{equation}\label{ds2WdS}
ds^2_{\rm WAdS} = ds^2_{\rm dS}(M, J_{\rm dS}) + 2 H^2 \chi_\mu \chi_\nu dx^\mu dx^\nu\,,
\end{equation}
with
\begin{equation}
\chi = \frac{1}{\sqrt{M + J_{\rm dS}/\ell_{\rm dS}} } \left( \partial_{t_E} - \frac{1}{\ell_{\rm dS}} \partial_{\phi} \right)
\end{equation}
and
\begin{equation}\label{ds2dS}
ds^2_{\rm dS} = - \left(M + \frac{R^2}{\ell_{\rm dS}^2} \right) dt_E^2 - \frac{dR^2}{M + \frac{J_{\rm dS}^2}{4 R^2} + \frac{R^2}{\ell_{\rm dS}^2} } + J_{\rm dS} dt_E d\phi + R^2 d\phi^2 \,.
\end{equation}
The indices of $\chi$ are raised and lowered with the metric $ds^2_{\rm dS}$. 
This metric solves Einsteins equation with $\Lambda = 1/ \ell_{\rm dS}^2$. The ground state of this metric is recovered by taking $M \to -1 $ and $ J_{\rm dS} \to 0$, which can be recognized as Euclidean de Sitter space in static coordinates. It can be obtained from the Lorentzian Kerr-de Sitter solutions of \cite{Balasubramanian:2001nb} by analytically continuing time and $J_{\rm dS}$, 
as well as taking $M \to -M$.
The metric \eqref{ds2WdS} solves the TMG field equations \eqref{TMGeom2} with\footnote{The orientation of the epsilon-tensor is chosen as $\epsilon^{t_E r \varphi} = \frac{1}{\sqrt{- \mathrm{det} g}}$}
\begin{equation}
\mu  \ell_{\rm dS} = - 3 \sqrt{2H^2 -1} \,, \qquad \Lambda = \frac{3+2H^2}{3\ell_{\rm dS}}  \,.
\end{equation}
This means that for real $\mu$, we should consider $2H^2 > 1$. 

The coordinates of \eqref{ds2WdS} are related to those of the metric \eqref{compdet} by:
\begin{subequations}
\begin{align}
	t & = \sqrt{\frac{(2H^2-1)}{\ell_{\rm dS}} (J_{\rm dS} + \ell_{\rm dS} M ) } \; t_E \\
	\theta & = \frac{t_E}{\ell_{\rm dS}} + \phi \\
	r & = \frac{2R^2 - J_{\rm dS} \ell_{\rm dS}}{4 \sqrt{M+\frac{J_{\rm dS}}{\ell_{\rm dS}}}} \,,
\end{align}
\end{subequations}
with
\begin{align}
	\label{Hnu}
H^2 = \frac{3+3 \nu^2}{6 - 2 \nu^2} \,, &&  \ell_{\rm dS} = - \frac{2 \ell}{\sqrt{3-\nu^2}} \,, &&
\nu^2 = \frac{6H^2 - 3}{2H^2 + 3} \,, && \ell^2 = \frac{3 \ell_{\rm dS}^2}{3+ 2H^2} \,.
\end{align}
and the mass and angular momentum parameters are related as
\begin{equation}\label{mjrel}
m = \frac16 \sqrt{M - \frac{J_{\rm dS} \sqrt{3-\nu^2}}{2\ell} } \,, \qquad j = \frac{ \nu J_{\rm dS} }{4 \sqrt{3-\nu^2}} \,.
\end{equation}
Hence the vacuum values of $m$ and $j$ can be related to the vacuum of the undeformed metric \eqref{ds2dS}, $M=-1$ and $J_{\rm dS} =0$ giving $m^{\rm vac} = i/6 $ and $j^{\rm vac} = 0$.

Because $H^2 > 1/2$, we are working in the domain where
\begin{equation}
0 < \nu^2 \leq 3\,.
\end{equation}
with $\nu^2 \to 3$ being the warped flat limit.

We are now ready to take the warped flat limit in the quadratic ensemble. 
	To this end, we reconsider the metric \eqref{ds2WdS} and after expressing $H$ and $\ell_{\rm dS}$ in terms of $\nu$ and $\ell$ using \eqref{Hnu}, we expand around  $\nu^2 = 3$. 
		Since $ 0 < \nu^2 \leq 3$, we expand the metric in a small parameter $\epsilon$ defined as
\begin{equation}
\nu^2  = 3 - \epsilon\,,
\end{equation} 
with $\epsilon > 0$. We also rescale
\begin{equation}
t_E \to \sqrt{\epsilon} \, t \,, \qquad J_{\rm dS} \to \sqrt{\epsilon} \, J_{\rm wf} \,.
\end{equation}
Now one can readily take $\epsilon \to 0$ in the metric \eqref{ds2WdS}. This limit effectively sends both the de Sitter radius and the deformation parameter $H$ to infinity, while keeping fixed their ratio. The resulting metric is the warped flat quotient in the following coordinates:
\begin{align}\label{ds2wfbh}
ds^2 = & \, 12 M dt^2 - 12 \left( J_{\rm wf} + \frac{r^2}{\ell} \right) dt d\phi - \frac{dr^2}{M} \nonumber \\
& + \left(\frac{3 J_{\rm wf}^2}{M} + r^2 \left(1 + 6 \frac{J_{\rm wf}}{\ell M} \right) + \frac{3 r^4}{M\ell^2}  \right) d\phi^2 \,.
\end{align}
This metric solves the TMG equation of motion \eqref{TMGeom2} with
\begin{equation}
\mu \ell = 3 \sqrt{3} \,, \qquad \Lambda = \frac{1}{\ell^2} \,.
\end{equation}
The reason why we call this metric warped flat now is also clear. The metric can be thought of as a deformation of flat space by terms of order $1/\ell$. Indeed, the limit $\ell \to \infty$ of \eqref{ds2wfbh} gives
\begin{equation}\label{ds2flatMJ}
ds^2_{\rm flat} = 12 M dt^2 - 12 J_{\rm wf} dt d\phi - \frac{dr^2}{M}  + \left(r^2 + \frac{3 J_{\rm wf}^2}{M} \right) d\phi^2
\end{equation}
By choosing $M=-1$ and $J_{\rm wf} = 0$ this metric is Minkowski space in three dimensions. For these values of $M$ and $J_{\rm wf}$  the metric \eqref{ds2flatMJ} possesses six globally well-defined Killing vectors (as opposed to two for generic values of $M$ and $J_{\rm wf}$ and four for $M=-1$ and $J_{\rm wf} \neq 0$).

The metric \eqref{ds2wfbh} is related to the warped flat quotient \eqref{wfbhxi}, which we will reinstate here for convenience:
\begin{equation}\label{ds2wfpaper}
\frac{ds^2}{\ell^2} = \frac{d\rho^2}{24 \xi \rho} - 24 \xi \rho d\varphi^2 + 12 \left[ du  + (\rho + \xi + \omega) d\varphi\right]^2 \,.
\end{equation}
The coordinate transformation which brings \eqref{ds2wfbh} to \eqref{ds2wfpaper} is
\begin{equation}
t = \frac{u \ell}{12 \xi} \,, \qquad r^2 = - 24 \ell^2 \xi \rho\,, \qquad \phi = \varphi  \,.
\end{equation}
and the parameters $M$ and $J_{\rm wf}$ are related to $\xi$ and $\omega$ by
\begin{equation}
M = 144\, \xi^2 \,, \qquad J_{\rm wf} = - 24 \,\ell \xi (\xi + \omega) \,,
\end{equation}
or, conversely:
\begin{equation}
\xi = \frac{\sqrt{M}}{12} \,, \qquad \omega = - \frac{M \ell + 6 J_{\rm wf}}{12 \sqrt{M} \ell} \,.
\end{equation}
From this, we see that the  values of $\omega$ and $\xi$ corresponding to the vacuum values of the undeformed metric $M = - 1$ and $J_{\rm wf} = 0$ are given by:
\begin{equation}
\xi^{\rm vac} = \frac{i}{12} \,, \qquad \omega^{\rm vac} = - \frac{i}{12} \,.
\end{equation}
For these vacuum values the deformed metric \eqref{ds2wfbh} is real and is related to the global warped flat spacetime \eqref{eqmetricvacuum} by the analytic continuation and coordinate transformation:
\begin{align}
x & = \frac{r}{\ell} \cos \phi \,, \\
- i \tau & =  \frac{r}{\ell} \sin \phi \,, \\
- i y & = \frac{t}{\ell} - \frac{r^2}{4 \ell^2} \sin 2\phi \,.
\end{align}

\section{Asymptotic symmetries of \texorpdfstring{WAdS$_3$}{WAdS3}}\label{app:WAdS}
In this appendix we will derive the asymptotic symmetry algebra for warped AdS$_3$ metrics \eqref{compdet} with positive cosmological constant $\Lambda = 1/\ell^2$ and $\nu^2 \neq 3$, using the methods of section \ref{TMGfirstorder}. For our purposes, it is convenient to remove the state-dependent parameters $\mathfrak{m}$ and $\mathfrak{j}$ from the $g_{rr}$ component of \eqref{compdet}, at the expense of introducing them in the $g_{t\theta}$ component. This is achieved by the state-dependent coordinate transformation
\begin{align}
r = & \; \frac{\rho}{2(\nu^2 - 3)} + \frac{6 \ell^2 \mathfrak{m}}{\nu^2-3} - \frac{2\ell^3 \left( (\nu^2 -3) \mathfrak{j} - 9 \nu \ell \mathfrak{m}^2\right)}{\nu(\nu^2-3) \rho} \,. \label{eq:rdef}
\end{align}
We furthermore define, with a considerable amount of hindsight, the parameters $\cM$ and $\CL$ in terms of the following combinations of $\mathfrak{m} $ and $\mathfrak{j}$ (and assuming $5v^2 - 3 \neq 0$)
\begin{equation}\label{redef}
\CL = \frac{ - (5\nu^2 - 3) \mathfrak{j} + 9 \ell \nu \mathfrak{m}^2}{6 \nu^2} \,, \qquad \cM = \mathfrak{m}\,.
\end{equation}
The metric then becomes
\begin{align} \label{eq:gWBan}
& d s^2 = d t^2 + \frac{\ell^2}{(\nu^2-3)\rho^2} d \rho^2 + 2 g_{t\theta} d t d \theta\, +  g_{\theta\theta} d \theta^2 \,. \nonumber
\end{align}
with:
\begin{align}
g_{t\theta} & = - \frac{1}{\nu^2 - 3} \left(\frac{\nu \rho}{\ell} +12 \ell \nu \cM + \frac{24 \ell^2 \nu^2( (\nu^2 - 3) \CL + 6 \ell \nu \cM^2 )}{ (5 \nu^2 -3) \rho} \right)\\
g_{\theta\theta} & =  g_{t\theta}^2 - \frac{\left( - (5\nu^2 -3 ) \rho^2 + 24 \ell^3 \nu (\nu^2-3) \CL + (12 \ell^2 \nu \cM)^2 \right)^2}{4 \ell^2 (3- 5\nu^2)^2 (\nu^2 -3) \rho^2} \,.
\end{align}
At first sight this metric may look much more inconvenient than \eqref{compdet}, but the advantage we gained is that we can now replace $\CL$ and $\cM$ by arbitrary functions of $\theta$ and it will still be an exact solution of the TMG field equations \eqref{TMGeom}. %\footnote{For $2r^2 + \ell^2(\nu^2-3) \CL >0$ and $\nu^2 > 3$ we have again $\mu\ell = 3\nu$, but when $\nu^2 < 3$ \eqref{eq:gWBan} solves \eqref{TMGeom} with $\mu\ell = - 3 \nu$. }
From now onward we will suppose that $\cM$ and $\CL$ are, indeed, arbitrary function of $\theta$. The asymptotic Killing vectors which preserve the form of the metric \eqref{eq:gWBan} as  $\rho \to \infty$ are
\begin{align}\label{zetaWAdS}
\zeta = & \,\left( T(\theta) + \frac{4 \ell^3 \nu Y''(\theta)}{(\nu^2 -3) \rho}  +  \frac{24 \ell^5 \nu \cM(\theta) Y''(\theta) }{(\nu^2-3) \rho^2} \right) \partial_t - Y'(\theta) \rho \, \partial_\rho \\
&  + \left( Y(\theta) +  \frac{2\ell^4 Y''(\theta)}{\rho^2}\right) \partial_\theta + \CO(\frac{1}{\rho^3}) \,. \nonumber
\end{align}
Following the analysis of section \ref{TMGfirstorder}, we will compute the charges and asymptotic symmetry algebra in the first-order formulation of TMG. We parameterize the metrics \eqref{eq:gWBan} by a frame field
\begin{subequations}
	\label{eq:triad}
	\begin{align}
	e_t & = T_1\,, \\
	e_\rho & = - \frac{\ell}{\sqrt{\nu^2-3}\,\rho} T_2\,, \\
	e_{\theta} & = - \left(\frac{\rho + 6 \ell^2 \cM}{\ell \sqrt{\nu^2 -3}} + \frac{\sqrt{\nu^2 -3}}{2\nu} g_{t\theta}\right) T_0 + g_{t\theta} T_1
	\end{align}
\end{subequations}
where we are assuming that $\nu^2 > 3$ (similar expressions exist for when $\nu^2<3$).

The field equations for TMG in first-order form \eqref{fieldeq} can be used to solve for the spin-connection $\omega$ and the auxiliary field $f$. The gauge-like transformations \eqref{eq:wbh5} which preserve the form of the Chern-Simons-like fields $a^\ione = \{e, \omega, f\}$ are given by \eqref{xiKilvec} with $\zeta$ given by the asymptotic Killing vectors \eqref{zetaWAdS}, up to a subleading term in the $\omega$-component of $\chi^\ione$. The asymptotic charges \eqref{varboundary} are integrable and finite in the limit $\rho \to \infty$. They are given by
\begin{equation}
Q[T,Y] = \frac{1}{2\pi G} \oint d \theta \left( T(\theta)\cM(\theta) + Y(\theta)\CL(\theta) \right)\,,
\label{eq:charges}
\end{equation}
where the state-dependent functions $\cM$ and $\CL$ transform under the asymptotic symmetry transformations generated by \eqref{zetaWAdS} as
\begin{subequations}
	\label{eq:wbh7}
	\begin{align}
	\delta_\zeta \cM & = - \frac{(\nu^2 -3)}{12\ell \nu} T'(\theta) + (\cM(\theta) Y(\theta) )' \,, \\
	\delta_\zeta \CL & = T'(\theta) \cM(\theta) +  Y(\theta) \CL'(\theta) + 2 Y'(\theta)\CL(\theta) - \frac{\ell(5\nu^2 - 3)}{12 \nu (\nu^2 -3)} Y'''(\theta)\,.
	\end{align}
\end{subequations}
Using the transformation laws \eqref{eq:wbh7} it is not hard to see that the Fourier modes $L_m = Q[T=0,Y=e^{im\theta}]$ and $P_m = Q[T=e^{im\theta},Y=0]$ satisfy the commutator algebra:
\begin{subequations}
	\begin{align}
	\label{eq:ASA}
	[L_n,L_m] & = (n-m) L_{m+n} + \frac{c}{12} n^3 \delta_{m+n,0}\, \\
	[L_n, P_m] & = - m P_{m+n} \\
	[P_n,P_m] & = \frac{k}{2} n \delta_{m+n,0}\,,
	\end{align}
\end{subequations}
with:
\begin{equation}
c =  \frac{\ell( 5\nu^2 - 3)}{\nu G (\nu^2 -3)} \,, \qquad  \qquad k = -  \frac{\nu^2 -3}{6 \nu \ell G}\,,
\label{eq:ccWdS}
\end{equation}
consistent with the values obtained in \cite{Detournay:2012pc}, upon analytically continuing $\nu \to i \nu$ and $\ell \to i \ell$.

The exact Killing vectors of the metric \eqref{compdet} become globally well-defined if 
\begin{equation}
2  \sqrt{9 m^2 - \frac{(\nu^2 - 3)j}{\ell \nu}} = i\,.
\end{equation}
In the limit to warped flat $\nu^2 \to 3$, this translates to the vacuum value of $P_0$
\begin{equation}
P_0^{\rm vac} = q = \frac{i}{6 G}\,.
\end{equation}
For this value of $P_0^{\rm vac}$, the vacuum value of $L_{0}$ is well-defined in the limit $\nu^2 \to 3$, even though both $c$ and $1/k$ diverge:
\begin{equation}
L_0^{\rm vac} = - \frac{c}{24} + \frac{q^2}{k} = - \frac{\ell}{24 \nu G}\,.
\end{equation}
The vacuum values of $P_0^{\rm vac} $ and $L_0^{\rm vac}$ can also be recovered from the quadratic ensemble, where they are related to the vacuum values of the undeformed locally de Sitter metric through \eqref{mjrel} and \eqref{redef}

\section{Derivations of Warped Cardy Entropy formula}
\label{appdercardy}
This appendix is meant to provide a reference for Cardy formulas and modular properties of any given WCFT. We first consider the theory to be defined on a complex plane described by coordinates $z, w$, which will be treated as independent complex coordinates. On this plane $T(z)$ denotes the right-moving energy momentum tensor and $P(z)$ denotes a right moving $\hat{u}(1)$ Kac-Moody current, which generate coordinate transformations of the form
 \begin{equation}
 \label{a2eq3}
z = f(z')\, \qquad w = w' +g(z')\,.
\end{equation}
These transformations are generated by infinitesimal vector fields $l_n = - z^{n+1} \partial_z$ and $p_n = -z^n \partial_w$ fulfilling
\begin{align}
\label{a2eq0}
     [ l_m, l_n ] &=  (m-n) l_{m+n}\,, \nonumber \\
     [ l_m, p_n ] &= - n p_{m+n}\,,  \\
    [ p_m, p_n ] &= 0\,.  \nonumber
\end{align}
The algebra of charges on the plane 
\begin{equation}
\label{a2eq2}
	L_n =  \frac{1}{2 \pi i} \oint d z\, z^{n+1} T(z)\,, \quad P_n = - \frac{1}{2 \pi} \oint d z \, z^n P(z)\,.
\end{equation}
is given by \eqref{a2eq0} up to central extensions.
The algebra \eqref{a2eq0} admits three non-trivial central extensions
\begin{align}
\label{a2eq1}
     [ L_m, L_n ] &=  (m-n) L_{m+n} + \frac{c}{12} (m^3-m) \delta_{n+m, 0}\,, \nonumber \\
     [ L_m, P_n ] &= - n P_{m+n} - i \kappa (m^2 +m) \delta_{m+n, 0}\,, \\
    [ P_m, P_n ] &= \frac{k}{2} m \delta_{n+m, 0}\,. \nonumber
\end{align}
The commutation relations \eqref{a2eq1} imply
 infinitesimal
transformation laws,
\begin{subequations}
\label{infi}
\begin{align}
\delta_\epsilon T(z) &= - \frac{c}{12} \partial_z^3 \epsilon(z) - 2 \partial_z \epsilon(z) T(z) - \epsilon(z) \partial_z T(z)\,,\\
\delta_\gamma T(z) &= - \partial_z \gamma(z) P(z) + \kappa \partial_z^2 \gamma(z)\,,\\
\delta_\epsilon P(z) &= - \partial_z \epsilon(z) P(z) - \epsilon(z) \partial_z P(z) - \kappa \partial_z^2 \epsilon(z)\,,\\
\delta_\gamma P(z) &= \frac{k}{2} \partial_z \gamma(z)\,.
\end{align}
\end{subequations}
where
$\epsilon(z')$ and $\gamma(z')$ are the infinitesimal
transformation parameters defined by
$z = z' - \epsilon(z')$
and $w = w' - \gamma(z')$ and
\begin{equation}
	\delta_\epsilon  = - i [T_\epsilon, \cdot]  \qquad
	\delta_\gamma  = - i [P_\gamma, \cdot]\,.
\end{equation}
Here, 
\begin{align}
    T_\epsilon &= -\frac{1}{2 \pi} \oint dz \epsilon(z) T(z)\,, &
    P_\gamma &= -\frac{1}{2\pi} \oint dz \gamma(z) P(z) \,.
\end{align}
The finite transformation laws may be inferred
by requiring that they reduce to the
infinitesimal versions \eqref{infi} and
 demanding consistency under composition \cite{Detournay:2012pc,Afshar:2015wjm}
\begin{subequations}
\label{a2eq4}
\begin{align}
T'(z') &= {\left( \frac{\partial z}{\partial z'} \right)}^2
\left( T(z) - \frac{c}{12} \{ z^\prime, z \} \right) + P(z) \frac{\partial w}{\partial z'} \frac{\partial z}{\partial z'}
- \kappa \left( \frac{\partial^2 w}{\partial^2 z'} - \frac{\partial w}{\partial z'} \frac{\partial z'}{\partial z} \frac{\partial^2 z}{\partial^2 z'} \right)\nonumber\\
&~ - \frac{k}{4} {\left( \frac{\partial w}{\partial z'} \right)}^2\\
P'(z') &= \frac{\partial z}{\partial z'} P(z) + \kappa \frac{\partial z'}{\partial z} \frac{\partial^2 z}{\partial^2 z'} - \frac{k}{2} \frac{\partial w}{\partial z'}\,.
\end{align}
\end{subequations}
In analogy to CFTs we use a warped conformal transformation
to map the theory to the plane
 \cite{Detournay:2012pc}
 \begin{equation}
   \label{a2eq22}
    z = \exp \left(i \varphi \right)\,, \qquad w = t + 2 \delta \varphi\,,
\end{equation}
where $\delta$ is a constant tilt parameter.

For now, we consider a WCFT on the cylinder with coordinates $(t, \varphi)$ chosen such that the symmetries are
\begin{equation}
    \label{a2eq5}
    \varphi^\prime = f(\varphi)\, \qquad
    t^\prime = t - g(\varphi)\,
\end{equation}
and put this theory on a circle $\varphi \sim \varphi + 2 \pi$.
The symmetries are generated by operators $T(\varphi)$
and $P(\varphi)$.
The coordinate identifications are
\begin{equation}
\label{a2eq6}
    \left(t, \varphi\right) \sim \left(t, \varphi + 2 \pi \right) \sim \left(t + i \beta, \varphi + \theta \right)\,.
\end{equation}
We will now use a transformation of the form \eqref{a2eq5} to exchange thermal and angular cycle
\begin{equation}
\label{a2eq7}
    \varphi^\prime = \lambda \varphi\, , \qquad t' = t - 2 \gamma \varphi\,,
\end{equation}
with
\begin{equation}
\label{a2eq8}
    \gamma = \frac{i \beta}{2 \theta}\, \qquad \lambda = \frac{2 \pi}{\theta}\,.
\end{equation}
This yields
\begin{equation}
\label{a2eq9}
    (t', \varphi') \sim \left(t' - i \beta', \varphi' - \theta' \right)\sim \left(t', \varphi' + 2 \pi \right)\, ,
\end{equation}
where
\begin{equation}
\label{a2eq10}
\beta' = \frac{2 \pi \beta}{\theta}\, \qquad \theta' = - \frac{4 \pi^2}{\theta}\,.
\end{equation}
The transformation behaviour of $T(\varphi), P(\varphi)$ under the transformation \eqref{a2eq7}
reads
\begin{subequations}
\begin{align}
    \label{a2eq11}
    P'(\varphi') &=\frac{1}{\lambda} P (\varphi) - k \frac{\gamma}{\lambda} \,,\\
    T'(\varphi') &= \frac{1}{\lambda^2} \left( T(\varphi)+ 2 \gamma P(\varphi) - \gamma^2 k\right)\,.
\end{align}
\end{subequations}
Using the Fourier decompositions
\begin{align}
\label{a2eq12}
	P_n &= - \frac{1}{2 \pi} \int_0^{2\pi} d \varphi\, e^{i n \varphi} P(\varphi)\,, \\
	L_n &= - \frac{1}{2 \pi} \int_0^{2 \pi} d \varphi\, e^{i n \varphi} T(\varphi)\,, \\
P_n' &= - \frac{1}{2 \pi} \int_0^{2 \pi \lambda} d \varphi' e^{i n \varphi'/\lambda} P(\varphi') \,, \\
L_n' &= - \frac{1}{2 \pi} \int_{0}^{2 \pi \lambda} d \varphi' e^{i n \varphi'/\lambda} T(\varphi') \,,
\end{align}
this implies the following transformation behaviour for the modes
\begin{align}
    \label{a2eq13}
    P_n^\prime &= P_n + \frac{i \beta}{2 \theta} k \delta_{n, 0}\\
    L_n^\prime &= \frac{\theta}{2 \pi} L_n + \frac{i \beta}{2 \pi} P_n - \frac{\beta^2}{8 \pi \theta } k \delta_{n, 0}\,.
\end{align}
The partition function at inverse temperature $\beta$ and
angular potential $\theta = i \beta \Omega$ transforms as
\begin{align}
\label{a2eq14}
    Z_{(0, 2 \pi)}(\beta, \theta) &= \mathrm{Tr}_{\{0, 2 \pi\}}\exp\left(-\beta P_0 + i \theta L_0 \right) =
    \mathrm{Tr}_{\{- \beta', -\theta'\}}\exp\left(2 \pi i L_0^\prime + \frac{i \beta^2}{4 \theta} k \right) \nonumber \\
    &= Z_{(- \beta', -\theta')}(0, 2 \pi) \exp \left(\frac{i \beta^2}{4 \theta} k \right)\,.
 \end{align}
This transformation can now be undone by a modular transformation (S-transformation). This acts on the torus, defined through the identifications
\begin{equation}
 z = i t + \varphi \sim z + n \alpha_1 + m \alpha_2\, ,
\end{equation}
with $\alpha_1 = 2 \pi$ and $\alpha_2 = i \beta + \theta$ as
\begin{equation}
S : (\alpha_1, \alpha_2) \rightarrow (- \alpha_2, \alpha_1)
\end{equation}
and therefore
 \begin{align}
 Z_{(0, 2 \pi)}(\beta, \theta)&= \exp \left( \frac{i \beta^2}{4 \theta} k  \right)\mathrm{Tr}_{\{0, 2 \pi\}}\exp\left(-\beta' P_0' + i \theta' L_0' \right)\nonumber \\
    &= \exp \left(\frac{i \beta^2}{4 \theta} k  \right)
    \mathrm{Tr}_{\{0, 2 \pi\}}\exp\left(- \frac{2 \pi \beta}{\theta} P_0^\prime - i \frac{4 \pi^2}{\theta} L_0^\prime \right)\,.
\end{align}
In the following, we drop the primes for $L_0'$ and $P_0'$, as the spectrum of the primed operators coincides with the original spectrum \cite{Detournay:2012pc} .
For $L_0$ bounded from below we find that, provided that
the vacuum is gapped, in the regime of small imaginary $\theta$
\begin{equation}
\label{a2eq15}
    Z(\beta, \theta) \sim \exp \left(\frac{i \beta^2}{4 \theta} k  \right)\exp\left(- \frac{2 \pi \beta}{\theta} P_0^{\mathrm{vac}} - i \frac{4 \pi^2}{\theta} L_0^{\mathrm{vac}} \right)\,.
\end{equation}
Computing the entropy via
\begin{equation}
    \label{a2eq16}
    S(P_0, L_0) = (1- \beta \partial_\beta - \theta \partial_\theta) \ln(Z)
\end{equation}
gives
\begin{equation}
    \label{a2eq17}
    S = - \frac{2\pi} {\theta}  \left( \beta P_0^{\mathrm{vac}} + 4 \pi i L_0^{\mathrm{vac}}\right)\,,
\end{equation}
where we still have to perform the Legendre transformation as $S = S(P_0, L_0)$. Here, we differentiate between two cases:\\
\textbf{1.) $k = 0$}
\begin{align}
\label{a2eq18}
    L_0^{k = 0} &= -i \frac{\partial \ln(Z)}{\partial \theta} = \frac{2\pi}{\theta^2} \left(2 \pi L_0^{\mathrm{vac}} - i \beta P_0^{\mathrm{vac}}  \right) \,, \\
    P_0^{k = 0} &= - \frac{\partial \ln(Z)}{\partial \beta} =  \frac{2 \pi}{\theta} P_0^{\mathrm{vac}} \,.
\end{align}
Expressing $\beta$ and $\theta$ in terms of $L_0$ and $P_0$ gives
\begin{equation}
    \label{a2eq19}
    S^{k = 0} = -\frac{2 i \pi  \left(L_0 {P_0^\mathrm{vac}}^2+ L_0^\mathrm{vac} P_0^2\right)}{P_0 P_0^\mathrm{vac}}\,.
\end{equation}
\textbf{2.) $k \neq 0$}
\begin{align}
\label{a2eq20}
    L_0^{k \neq 0} &= \frac{4\pi^2}{\theta^2} L_0^{\rm vac} - \frac{\beta}{4 \theta^2}  (\beta  k + 8\pi  i   P_0^\mathrm{vac}) \,, \\
    P_0^{k \neq 0} &= - \frac{\partial \ln(Z)}{\partial \beta} =  \frac{2 \pi}{\theta}  P_0^{\mathrm{vac}} - \frac{i k \beta}{2 \theta}\,.
\end{align}
Expressing $\beta$ and $\theta$ in terms of $L_0$ and $P_0$ gives 
\begin{align}
\label{a2eq21}
S = - \frac{4 i P_0 P_0^{\mathrm{vac}} \pi}{k} + 4 \pi \sqrt{- \left( L_0 - \frac{P_0^2}{k} \right) \left( L_0^\mathrm{vac} - \frac{{P_0^\mathrm{vac}}^2}{k}
\right)
} \,.
\end{align}

Both \eqref{a2eq19} and \eqref{a2eq21} depend on the vacuum
of our theory. One common way to define the vacuum state is
to define it as the state of maximal symmetry. As,
$L_0$ and $P_0$ generate global symmetries its action on
the vacuum should give 0. However, one cannot
choose the vacuum expectation values to be zero on the plane
and on the cylinder,
since under the map \eqref{a2eq22} $T(z)$ and $P(z)$
transform as
\begin{subequations}
\begin{align}
    \label{a2eq23}
    T'(\varphi) &= - z^2 T(z) + 2 i P z \delta + \left(\frac{c}{24} + 2 i \kappa \delta - k \delta^2 \right)\,,\\
    P'(\varphi) &= i z P(z) + i \kappa - \delta k\,.
\end{align}
\end{subequations}
Using the decompositions into Fourier and Laurent modes
this implies
\begin{subequations}
\begin{align}
\label{a2eq25}
    L_n^{\mathrm{cyl}} &= L_n^{\mathrm{plane}} + 2 \delta P_n^{\mathrm{plane}} - \left( 2 i \kappa \delta + \frac{c}{24} - k \delta^2 \right) \delta_{n, 0}\,, \\
    P_n^{\mathrm{cyl}} &= P_n^{\mathrm{plane}} - \left(i \kappa -  k \delta\right) \delta_{n, 0}\,.
\end{align}
\end{subequations}
Choosing $L_0^{\mathrm{plane}} = P_0^{\mathrm{plane}} = 0$
on the plane where the algebra was initially defined,
we obtain
\begin{subequations}
	\label{a2eq26}
\begin{align}
    L_0^{\mathrm{cyl, vac}} &=  - 2 i \kappa \delta - \frac{c}{24} + \delta^2 k\,, \\
    P_0^{\mathrm{cyl, vac}} &= - i \kappa +  \delta k \,.
\end{align}
\end{subequations}

\bibliographystyle{utphys}
\bibliography{warpedflat}

\end{document}